\documentclass[aps,nofootinbib,preprintnumbers,superscriptaddress]{revtex4}
\usepackage{epsfig}
\usepackage{amssymb,amsmath,bm,bbm}
\usepackage{graphicx}
\usepackage{verbatim}
\usepackage{appendix}
\usepackage[usenames,dvipsnames]{xcolor}

\begin{document}

\preprint{YITP-16-142, IPMU-16-0206}

\title{Stable solutions of inflation driven by vector fields}

\author{Razieh Emami}
\affiliation{Institute for Advanced Study, The Hong Kong University of Science and Technology, Clear Water Bay, Kowloon, Hong Kong}

\author{Shinji Mukohyama}
\affiliation{Center for Gravitational Physics, Yukawa Institute for Theoretical Physics, Kyoto University, 606-8502, Kyoto, Japan}
\affiliation{Kavli Institute for the Physics and Mathematics of the Universe (WPI), The University of Tokyo Institutes for Advanced Study, The University of Tokyo, Kashiwa, Chiba 277-8583, Japan}

\author{Ryo Namba}
\affiliation{Department of Physics, McGill University, Montr\'{e}al, QC, H3A 2T8, Canada}
\affiliation{Kavli Institute for the Physics and Mathematics of the Universe (WPI), The University of Tokyo Institutes for Advanced Study, The University of Tokyo, Kashiwa, Chiba 277-8583, Japan}

\author{Ying-li Zhang}
\affiliation{National Astronomy Observatories, Chinese Academy of Science, Beijing 100012, People's Republic of China}
\affiliation{Institute of Cosmology and Gravitation, University of Portsmouth, Portsmouth PO1 3FX, UK}

\begin{abstract}
Many models of inflation driven by vector fields alone have been known to be plagued by pathological behaviors, namely ghost and/or gradient instabilities. In this work, we seek a new class of vector-driven inflationary models that evade all of the mentioned instabilities. We build our analysis on the Generalized Proca Theory with an extension to three vector fields to realize isotropic expansion. We obtain the conditions required for quasi de-Sitter solutions to be an attractor analogous to the standard slow-roll one and those for their stability at the level of linearized perturbations. Identifying the remedy to the existing unstable models, we provide a simple example and explicitly show its stability. This significantly broadens our knowledge on vector inflationary scenarios, reviving potential phenomenological interests for this class of models.
\end{abstract}

\maketitle

\section{Introduction}

Inflation has since its birth been a successful paradigm to resolve the horizon, flatness and unwanted relics problems in the Hot Big Bang cosmology as well as providing the initial seeds for fluctuations in the cosmic microwave background (CMB) radiation and for the large-scale structure formation.%
\footnote{Some alternative mechanisms to inflation have been proposed. A small subset of them consists of
pre-big-bang scenario \cite{Gasperini:2002bn} (and references therein), string gas cosmology \cite{Brandenberger:1988aj}, matter bounce \cite{Wands:1998yp,Finelli:2001sr}, Ekpyrotic scenario \cite{Khoury:2001wf,Buchbinder:2007ad}, cosmology in Ho\v{r}ava-Lifshitz gravity \cite{Horava:2009uw,Mukohyama:2009gg}, and Galilean  Genesis \cite{Creminelli:2010ba}.}
A number of past, ongoing and upcoming CMB experiments, space-based \cite{Bennett:1996ce} (latest results by Planck \cite{Ade:2015lrj}), balloon-borne \cite{Netterfield:2001yq}, and ground-based \cite{Carlstrom:2009um}, as well as those for the large-scale structure (LSS), aim to measure and/or constrain inflationary parameters. The observational precisions of the forthcoming experiments are expected to be yet at unprecedented levels, and the nomenclature {\it precision cosmology} is more appropriate than has ever been.

Most of the realizations of inflation rely on slow roll of one or multiple scalar fields.%
\footnote{See, e.g., \cite{Martin:2013tda,Martin:2013nzq} for a good collection of models. There are examples that do not require the slow roll, such as $k$-inflation \cite{ArmendarizPicon:1999rj}, where the inflationary solution is supported by the kinetic term of a scalar field.}
The slow roll is to ensure a prolonged period of inflationary stage, and the use of scalar fields is the simplest realization in field theories that is consistent with a spatially isotropic expansion.
On the other hand, possibilities to obtain inflationary solutions driven by higher-spin fields, especially vector fields, have been sought for.%
\footnote{There are examples of inflation by $p$-form fields in \cite{Kaloper:1991rw}.}
Since the standard $U(1)$ gauge field only with the kinetic action $- F^2 /4$ is conformally coupled to gravity and thus its energy density quickly decays away, the attempts of vector-driven inflation often employed breaking of gauge invariance, such as non-minimal coupling to gravity \cite{Turner:1987bw,Golovnev:2008cf,Kanno:2008gn,Dimopoulos:2008rf}, introduction of potential terms $V(A^2)$ \cite{Ford:1989me,Koivisto:2008xf}, and fixing the norm by Lagrange multiplier \cite{Ackerman:2007nb}.
However, these models contain a longitudinal mode of the vector field that has been found to suffer instabilities on inflationary backgrounds, and therefore all have turned out to be pathological \cite{Himmetoglu:2008hx,Himmetoglu:2008zp,Himmetoglu:2009qi}.

A different model was proposed in \cite{Watanabe:2009ct}, in which the $U(1)$ gauge invariance is preserved while the kinetic term of the vector (gauge) field is modified by a scalar field $\phi$ with a flat potential, through a $- I^2(\phi) F^2$ term, and this coupling leads the vector to an attractor phase and prevents it from decaying away by expansion. While the dominant energy component is the one of $\phi$ and therefore this is not a vector-driven inflation {\it per se}, the dangerous longitudinal mode is absent in this model. The same type of coupling but without the vector vacuum expectation values (vev) has been extensively studied in the context of primordial magnetogenesis \cite{Ratra:1991bn,Bamba:2003av,Martin:2007ue,Demozzi:2009fu,Emami:2009vd,Caldwell:2011ra,Fujita:2012rb,Ferreira:2013sqa,Kobayashi:2014sga,Ferreira:2014mwa,Domenech:2015zzi,Campanelli:2015jfa,Fujita:2016qab}, while it has been shown that the CMB constraints on higher-order correlation statistics place stringent bounds on the produced amplitude of magnetic fields \cite{Barnaby:2012tk,Fujita:2013pgp,Fujita:2014sna}.
This is the only coupling between a scalar and a vector field that respects gauge invariance and parity, without invoking derivatives.%
\footnote{Some derivative coupling operators and their effects on magnetogenesis have been considered in  \cite{Tasinato:2014fia,Mukohyama:2016npi}.}
In the case of a pseudo-scalar $\phi$, symmetries admit a coupling to a gauge field in the form $\phi F \tilde{F}$, where $\tilde F$ is the dual of the field-strength tensor $F$. It results in a copious, exponential production of the vector field, and its backreaction can slow down the motion of $\phi$ that would otherwise roll down its potential quickly \cite{Anber:2009ua}.
In addition, this model turns out to have rich
phenomenological signatures such as non-Gaussianity \cite{Barnaby:2010vf,Barnaby:2011vw}, helical gravitational waves \cite{Cook:2011hg,Barnaby:2011qe,Crowder:2012ik,Namba:2015gja,Obata:2016oym,Maleknejad:2016qjz,Obata:2016tmo}, magnetogenesis \cite{Caprini:2014mja,Fujita:2015iga,Adshead:2016iae}, baryogenesis \cite{Anber:2015yca,Cado:2016kdp}, and primordial black holes \cite{Linde:2012bt,McDonough:2016xvu} (for further review, see \cite{Pajer:2013fsa}).%
\footnote{The same type of coupling can invoke a rotation of photon polarizations in the propagation between the last scattering surface and the present time, called cosmological birefringence \cite{Lue:1998mq}.}

Ref.~\cite{Adshead:2012kp} promoted the Abelian vector to a non-Abelian gauge field coupled to a pseudo-scalar $\phi$ as in the Abelian case, dubbing the model as Chromo-natural inflation. Another model called Gauge-flation proposed in \cite{Maleknejad:2011jw,Maleknejad:2011sq}, which can be interpreted as a limit of Chromo-natural model with $\phi$ heavy and integrated out \cite{SheikhJabbari:2012qf,Adshead:2012qe}, was a novel vector-driven model of inflation. These models employ an $SU(2)$ gauge field, and due to the local property $SU(2) \cong SO(3)$, one can orient the vev of the three vector fields along the three spatial directions, and then an isotropic expansion is manifestly an (local) attractor \cite{Maleknejad:2011jr}.
Unfortunately, however, neither of these models survives against observational constraints: essentially the tensor modes experience a tachyonic growth for a limited duration around horizon crossing, and the tensor-to-scalar ratio is always beyond the level of observational upper bound, for any acceptable values of the scalar spectral index \cite{Dimastrogiovanni:2012ew,Adshead:2013nka,Namba:2013kia}.
Recently, the variants of these models, with massive $SU(2)$ fields, were considered, dubbed Massive Gauge-flation \cite{Nieto:2016gnp} and Higgsed Chromo-natural \cite{Adshead:2016omu}, and it has been proposed that the mass terms can enlarge the parameter space so that the model predictions are in agreement with the data.

As described above, among the plethora of the models, almost all of the vector-only-driven ones are either theoretically inconsistent due to ghost/gradient instabilities or observationally unfavored. So far, the only known stable models of inflation solely driven by vector fields are the Gauge-flation and its massive variant, and they involve the $SU(2)$ structure constants $\epsilon_{abc}$ in an essential way through the term of the form $\epsilon_{abc}\partial_{[\mu}A^{(a)}_{\nu]}A^{(b)\mu}A^{(c)\nu}$. However, as mentioned above, the Gauge-flation with the unbroken $SU(2)$ gauge symmetry is observationally disfavored. Once the $SU(2)$ gauge symmetry is abandoned, the $SU(2)$ specific structure is not necessary and a more general class of models is allowed/motivated. (The $SU(2)$ specific structure is kept if the gauge symmetry is broken spontaneously but in this case the system inevitably involves an additional scalar field, i.e.~a Higgs field.) Therefore, our aim of this paper is, as a first step towards a wider class of models of vector-only driven inflation, to search a stable inflationary solution driven by vector fields without the mentioned $SU(2)$ specific structure. This would significantly broaden our knowledge on vector-driven inflationary models.

Taking the lessons from the preceding unsuccessful cases, we start from a highly general class of vector-field models: Generalized Proca theory. This theory has been introduced in \cite{Tasinato:2014eka,Heisenberg:2014rta,Allys:2015sht,Jimenez:2016isa,Allys:2016jaq} and consists of a vector field without gauge invariance, and thus its longitudinal mode propagates as a physical degree of freedom. While it contains derivative terms of the vector field in the action without gauge invariance, the theory is constructed in such a way that it takes the most general form with which the variations of the action give rise only to the equations up to second-order derivatives. Therefore, Generalized Proca theory is by construction guaranteed to be free from higher-derivative instabilities, also known as Ostrogradsky ghosts \cite{Ostrogradsky:1850fid}.%
\footnote{Recently an inflationary solution is investigated in a model in which vector fields couple non-minimally to gravity \cite{Oliveros:2016myr}; however, this model is neither within the class of Generalized Proca theory nor contains the terms, identified in our work, to cure ghosts at the linearized order.}
The theory is quite general, as it is characterized by several free functions (five, in the most general case -- further extensions are proposed in \cite{Heisenberg:2016eld,Kimura:2016rzw}). In this work, we investigate this general class and identify the conditions under which all the perturbations around inflationary backgrounds are stable against both ghost and gradient instabilities. We also restrict our attention to the case where the background is quasi de-Sitter with an attractor behavior analogous to the standard slow-roll attractor.

Some deformation of the original theory is necessary for our current purpose.
In \cite{DeFelice:2016yws,DeFelice:2016uil}, the application of this theory to the late-time accelerated expansion was analyzed. There the vector field is given a vev in its temporal component, and for this reason, the value of the Hubble parameter (and the vev) is exactly constant, not allowing a quasi de-Sitter solution.
On the other hand, when its spatial components take a vev, a vector field, as its name stands, inevitably points to a specific direction in space. An expansion driven by such a field therefore has a privileged direction, and the correlation functions become statistically anisotropic \cite{Ackerman:2007nb}. In fact, the model in \cite{Watanabe:2009ct} was studied intensively in the context of statistical anisotropy \cite{Watanabe:2010fh,Emami:2010rm,Gumrukcuoglu:2010yc,Emami:2011yi,Bartolo:2012sd,Soda:2012zm,Shiraishi:2012xt,Emami:2013bk,Abolhasani:2013bpa,Lyth:2013kah,Baghram:2013lxa,Ohashi:2013qba,Biagetti:2013qqa,Shiraishi:2013vja,Ramazanov:2013wea,Abolhasani:2013zya,Chen:2014eua,Naruko:2014bxa,Abolhasani:2015cve,Emami:2015qjl,Emami:2015uva},%
\footnote{Vector curvaton scenarios in a resembling circumstance were considered in \cite{Dimopoulos:2008yv,Dimopoulos:2009am,Dimopoulos:2009vu,Dimopoulos:2011ws,Namba:2012gg}.}
and anisotropic solutions in Generalized Proca are explored in \cite{Heisenberg:2016wtr}.
However, since the amount of anisotropy is stringently constrained by the Planck data \cite{Kim:2013gka}, we deform the original theory so as to minimize the anisotropic configuration, while preserving the no-Ostrogradsky-ghost construction. To this end, we introduce $3$ vector fields and correspondingly give each of their spatial components a homogeneous vev pointing in each of the spatial directions perpendicular to each other, namely $\langle A^{(a)}_i \rangle = A(t) \delta^a_i$, where the superscript $(a)$ is the label of the three vectors. Moreover, in order to set up the theory consistent with this background configuration, we impose a global $O(3)$ symmetry in the field space and retrieve the terms that respect it.%
\footnote{During our preparation of this manuscript, generalization of Generalized Proca to multiple vectors appeared in the literature \cite{Jimenez:2016upj}. Our deformation of the original theory is mutually consistent with this work, under the concerned symmetry. }
This setup, together with the conditions for a background attractor, indeed ensures an isotropic expansion of the universe and vanishing statistical anisotropy, at least in the linearized perturbations.

We focus on quasi de-Sitter solutions for the background, i.e.~almost constant Hubble parameter $\dot H \approx 0$, that are given by a slowly varying physical vector vev, i.e.~$\partial_t (A / a) \approx 0$, where $a$ is the scale factor. Our interest in this work is to analyze the stability of these background solutions. Specifically, we investigate the model window that satisfies all of the following conditions: (i) the one for the quasi de-Sitter solutions to be an analogue of the standard slow-roll attractor, (ii) the one to preserve the background isotropic configuration and geometry, and (iii) the ones for the stabilities of perturbations against both ghost and gradient instabilities.
The condition (ii) is to ensure that the vector vev does not run away from $\langle A^{(a)}_i \rangle \propto \delta^a_i$ and therefore that no anisotropic part of the background metric is excited dynamically.
The perturbations are decoupled into $3$ sectors at the linearized order, analogous to the standard tensor/vector/scalar decomposition,%
\footnote{This analogy is only an approximate one, since the decomposition of the triplet vector fields is not the standard one under spatial rotation. See Sec.~\ref{sec:perturbations} and refs.~\cite{Maleknejad:2011jw,Dimastrogiovanni:2012ew} for clarification.}
and we separately examine the stability conditions for each sector.
Our analysis clarifies that the terms of the form $A_\mu A_\nu F^{\mu\rho} F^\nu{}_\rho$ are crucial for the absence of ghosts, and no inflationary models of vector fields can be stable without those terms within the class of model considered in the present paper. (See (\ref{action})-(\ref{def-XYZW}) below.)
The conditions for gradient stability are quite involved, and those in the general case are not illuminating. We therefore consider some examples, and then provide a (relatively) simple example that can indeed satisfy all the conditions simultaneously.
This is the first example of inflationary model in which vector fields alone, without the structure specific to the $SU(2)$ gauge symmetry, drive quasi de Sitter inflationary expansion.

This paper is organized as follows. In Sec.~\ref{sec:setup}, we introduce the theory we consider and describe the necessary deformation of the original Generalized Proca theory to have isotropic expansion. In Sec.~\ref{sec:background}, we seek for quasi de-Sitter background solutions and in particular provide the attractor condition. Sec.~\ref{sec:perturbations} is devoted for the stability analyses of perturbations: we outline the general procedure in Sec.~\ref{subsec:general}, then proceed the analysis for tensor, vector and scalar sectors in Secs.~\ref{subsec:tensor}, \ref{subsec:vector} and \ref{subsec:scalar}, respectively,
and provide the condition to ensure the background attractor to be isotropic in Sec.~\ref{subsec:isotropy}. We summarize all the conditions in Sec.~\ref{subsec:stability-all}. Sec.~\ref{sec:examples} illustrates some examples: we consider existing (therefore unstable) models and identify the instabilities in Sec.~\ref{subsec:example1} and then provide a successful model with all the stability conditions satisfied in Sec.~\ref{subsec:example2}. We finally conclude in Sec.~\ref{sec:conclusion}.
Throughout the paper we use the natural units with $\hbar = c = 1$, take the $\left( - + + + \right)$ metric signature, and denote the reduced Planck mass by $M_p$.

\section{Generalized Proca Theory with $O(3)$ Invariance}
\label{sec:setup}

Our aims are to search for a model of inflation that is driven solely by vector fields, to identify necessary building blocks and to broaden the class of allowed models. It has been known that simple implementations of this are plagued by instabilities \cite{Himmetoglu:2008hx,Himmetoglu:2008zp,Himmetoglu:2009qi} and thus we need to resort to more dedicated models. In this regard, we turn our attention to the so-called generalized Proca theory, introduced in \cite{Tasinato:2014eka,Heisenberg:2014rta,Allys:2015sht,Jimenez:2016isa,Jimenez:2016upj,Allys:2016jaq} as the straightforward extension of the Horndeski scalar-tensor theory \cite{Horndeski:1974wa,Nicolis:2008in,Deffayet:2011gz,Deffayet:2009wt,Kobayashi:2011nu} to a vector-tensor one (thus free from Ostrogradsky instabilities \cite{Ostrogradsky:1850fid}).
As already stated in the introduction, a single vector field with a timelike vev does not allow for a graceful exit from inflation. On the other hand, a single spacelike vector field intrinsically carries a privileged direction and hence inevitably introduces anisotropy. In order to realize inflation with isotropic expansion, we thus introduce three vector fields $A^{(1,2,3)}_\mu$ with vacuum expectation values (vev),%
\footnote{One can start from $\langle A^{(a)}_0 \rangle \ne 0$ instead, but then they can see that $\langle A^{(a)}_0 \rangle = 0$ is a trivial solution of the background. We thus set this from the beginning.}
\begin{equation}
\langle A_i^{(a)} \rangle = A(t) \, \delta^a_i \; , \quad
\langle A_0^{(a)} \rangle = 0 \; .
\label{A-vev}
\end{equation}
Moreover, in order to preserve this background configuration, we impose a global $O(3)$ symmetry in the internal field space.%
\footnote{In the case of $SU(2)$ gauge fields, such as in Chromo-natural inflation \cite{Adshead:2012kp} and gauge-flation \cite{Maleknejad:2011jw}, or any larger group that has $SU(2)$ as its subgroup, this type of symmetry is naturally realized since $SU(2)$ is homomorphic to $SO(3)$.}
Requiring the form of Generalized Proca action \cite{Heisenberg:2014rta,Allys:2015sht,Jimenez:2016isa,Jimenez:2016upj,Allys:2016jaq} with three vectors and the symmetry and restricting to a class of simple actions, we have
\begin{equation}
S = \int d^4x \sqrt{-g} \left( L_2 + L_4 \right) \; ,
\label{action}
\end{equation}
where~%
\footnote{One can also include in $G_2$ the terms $A_\mu A_\nu \tilde{F}^{\mu\rho} \tilde{F}^{\nu}{}_\rho$, but they can be re-expressed in terms of $X$, $Y$ and $W$ and therefore are redundant.}
\footnote{Here we set parameters $c_2$ in the $L_4$ term \cite{Heisenberg:2014rta} to be zero for simplicity. Since they vanish from the action in the scalar limit $A_\mu \rightarrow \partial_\mu \pi$, they do not contribute to the background dynamics and affect only the vector (and tensor in our calculations) sectors \cite{DeFelice:2016yws}. Recovering them can be done straightforwardly, and the procedure of our subsequent calculations will be the same.}
\begin{eqnarray}
L_2 &=& G_2 (X,Y,Z,W_1,W_2,W_3) \; , \quad
L_4 = G_4(X) \, R + G_{4,X} \sum_{a=1}^3\left[ \nabla_\mu A^{(a) \, \mu} \nabla_\nu A^{(a) \, \nu} - \nabla_\mu A^{(a) \, \nu} \nabla_\nu A^{(a) \, \mu} \right] \; ,
\label{def-Ls}\\
&& \!\!\!\!\! \!\!\!\!\!
\begin{aligned}
& X \equiv - \frac{1}{2} \sum_{a=1}^3 A_\mu^{(a)} A^{(a)\, \mu} \; , \quad
Y \equiv - \frac{1}{4} \sum_{a=1}^3 F_{\mu\nu}^{(a)} F^{(a) \, \mu\nu} \; , \quad
Z \equiv - \frac{1}{4} \sum_{a=1}^3 F_{\mu\nu}^{(a)} \tilde F^{(a) \, \mu\nu} \; , \\
& W_1 \equiv \sum_{a,b=1}^3 A_\mu^{(a)} A_\nu^{(a)} F^{(b) \, \mu\rho} \, F^{(b) \, \nu}{}_\rho \; , \quad
W_2  \equiv \sum_{a,b=1}^3 A_\mu^{(a)} A_\nu^{(b)} F^{(a) \, \mu\rho} \, F^{(b) \, \nu}{}_\rho \; , \quad
W_3 \equiv \sum_{a,b=1}^3 A_\mu^{(a)} A_\nu^{(b)} F^{(b) \, \mu\rho} \, F^{(a) \, \nu}{}_\rho \; ,
\end{aligned}
\label{def-XYZW}
\end{eqnarray}
with $F^{(a)}_{\mu\nu} = \nabla_\mu A^{(a)}_\nu - \nabla_\nu A^{(a)}_\mu$, $\tilde F^{(a) \, \mu\nu} = \epsilon^{\mu\nu\rho\sigma} F^{(a)}_{\rho\sigma} / 2$, and $G_{4,X} \equiv \partial G_4 / \partial X$.
Notice the different contractions in the definitions of $W_{1,2,3}$, and that the superscript on the vectors is merely a label of the three vector fields. Also we have excluded the $L_6$ term for simplicity.
Although there are in principle many other possible ways to contract the $(a)$ indices, such as $A_\mu^{(a)} A^{(b) \, \mu} A_\nu^{(a)} A^{(b) \, \nu}$ etc., we take the minimal choice as in \eqref{def-XYZW} and demonstrate that it equips sufficient room to provide stable inflationary solutions.
More importantly, we exclude the dependence of $G_2$ on yet another $SO(3)$ ($\simeq SU(2)$) invariant scalar combination $\epsilon_{abc}F^{(a)}_{\mu\nu}A^{(b)\mu}A^{(c)\nu}$, where $\epsilon_{abc}$ represents the $SO(3)$ ($\simeq SU(2)$) structure constants. This specific combination is included in known models of vector-only driven inflation, i.e. Gauge-flation and its massive variant. In the present paper we shall find a stable model of inflation without relying on the combination $\epsilon_{abc}F^{(a)}_{\mu\nu}A^{(b)\mu}A^{(c)\nu}$ and thus significantly broaden a class of allowed models of inflationary models driven solely by vector fields.
Note that the case of $G_2 = Y$ and $G_4 = 0$ is the standard free $U(1)$ gauge theory, that of $G_2 = Y + m^2 X$ ($G_4 = 0$) the original Proca theory with mass $m$, and that of $G_2 = Y + m^2 X$ and $G_4 = M_p^2 / 2 - X / 6$ the non-minimally coupled model considered in e.g.~\cite{Golovnev:2008cf,Kanno:2008gn,Dimopoulos:2008rf} (with the terms proportional to $G_{4,X}$ dropped by hand). Eq.~\eqref{action} is the action we study in this work, and we derive the stability conditions around the inflationary background.

\section{Inflationary Background}
\label{sec:background}

The class of models given by \eqref{action} consists of three vector fields $A^{(a)}_\mu$, and their vev \eqref{A-vev} drives inflation. We assume the background geometry to be the flat Friedmann-Lema\^{i}tre-Robertson-Walker (FLRW) metric,
\begin{equation}
ds^2 = - dt^2 + a^2(t) \delta_{ij} dx^i dx^j \; ,
\label{FLRW}
\end{equation}
being consistent with the vector vev configuration. We then have
\begin{equation}
\bar X = - \frac{3A^2}{2 \,  a^2} \; , \quad
\bar Y = \frac{3 \dot A^2}{2 \, a^2} \; , \quad
\bar Z = 0 \; , \quad
\bar W_1 = \frac{\bar W_2}{3} = \bar W_3 = - \frac{3 A^2 \dot A^2}{a^4} \; ,
\label{XYZW-back}
\end{equation}
where bar denotes background quantities, and dot denotes derivative with respect to physical time $t$. $Z$ vanishes on the background, since $Z \propto \epsilon^{ijk} \partial_t A^{(a)}_i \partial_j A^{(a)}_k$ and the homogeneous vector vev does not survive against the spatial derivative on it.
The background action reads
\begin{equation}
S^{(0)} = V \int dt \, a^3 \left[ \bar{G}_2 - 6 \, \bar G_4 \, \frac{\dot{a}^2}{a^2} + 12  \, \bar G_{4,X} \, \frac{\dot a}{a} \, \frac{A}{a} \, \partial_t \left(\frac{A}{a} \right)  \right]
\end{equation}
where $V$ is the comoving volume.
There are two dynamical degrees of freedom in the background, namely $A(t)$ and $a(t)$. The corresponding equations of motion are, respectively,
\begin{eqnarray}
&&
\begin{aligned}
& \partial_t \left\{ \left[
\bar G_{2,Y}  - 2 \left( \bar G_{2,W_1} + 3 \bar G_{2,W_2} + \bar G_{2,W_3} \right) B^2 \right] \left( \dot B + HB \right) \right\}
+ 2H \bar G_{2,Y} \left( \dot B + HB \right)
+ \bar G_{2,X} B
\\ & \qquad\quad
+ 2 \left( \bar G_{2,W_1} + 3 \bar G_{2,W_2} + \bar G_{2,W_3} \right) B
\left( \dot B^2 - H^2 B^2 \right)
+ 2 \bar G_{4,X} B \left( 2 \dot H + 3 H^2\right) = 0
\; ,
\end{aligned}
\label{EQB} \\
&&
\begin{aligned}
& 4 \bar G_4 \dot H
+ 2 \left[ \bar G_{2,Y} - \left( \bar G_{2,W1} + 3 \bar G_{2,W2} + \bar G_{2,W3} \right) B^2 \right] \left( \dot B + HB \right)^2
+ \bar G_{2,X} B^2
\\ & \qquad\quad
- 4 \, \partial_t \left( \bar G_{4,X} B \dot B \right)
+ 2 \bar G_{4,X} B^2 \left( 2 \dot H + 3 H^2 \right)
= 0
\; .
\end{aligned}
\label{EQHdot}
\end{eqnarray}
together with the constraint equation
\begin{equation}
6 \bar G_4 H^2 = - \bar G_2 + 12 \, \bar G_{4,X} HB \dot B
+ 3 \left[ \bar G_{2,Y} - 2 \left( \bar G_{2,W1} + 3 \bar G_{2,W2} + \bar G_{2,W3} \right) B^2 \right] \left( \dot B + HB \right)^2 \; ,
\label{EQconst}
\end{equation}
where we have defined the Hubble parameter $H$ and ``physical'' vector vev $B$ as \footnote{The word "physical" means that the value of $B$ is a tetrad component of the vectors and thus does not depend on the normalization of the scale factor $a$.}
\begin{equation}
H \equiv \frac{\dot a}{a} \; , \quad
B \equiv \frac{A}{a} \; .
\end{equation}
Note that the choice of $G_2 = Y + m^2 X$ and $G_4 = M_p^2 /2$ correctly reproduces the equations of motion and of state for the Proca theory minimally coupled to gravity.

In order to obtain an inflationary solution, we specifically look for one with quasi de Sitter expansion, namely,
\begin{equation}
\dot H \approx 0 \; .
\label{dS-H}
\end{equation}
One can realize that background equations have \eqref{dS-H} as a solution only if the vector vev $B$ is in the slow-roll phase, i.e.,
\begin{equation}
\dot B \approx 0 \; .
\label{dS-B}
\end{equation}
This can be seen by setting $\dot{H} = 0$ in the background equations of motion. First, \eqref{EQconst} can be solved algebraically with respect to $\dot{B}$. Hence, $\dot{B}$ and $\ddot{B}$ can be expressed as functions of $H$ and $B$, where we set $\dot{H}=0$ to obtain the expression for $\ddot{B}$. By substituting them to \eqref{EQHdot} and setting $\dot{H}=0$, we obtain an algebraic equation involving $H$ and $B$ only. Solving this results in $B=B(H)$.
Therefore, barring fine-tunings, given a constant $H$ as in \eqref{dS-H}, the background equations only admit a constant $B$ as an inflationary solution. Moreover, once $\dot H$ in \eqref{EQB} is replaced with the use of \eqref{EQHdot}, one observes that \eqref{EQB} can be written as $\ddot B = h(H,B,\dot{B}) \dot B$, where $h(H,B,\dot{B})$ is non-singular at $\dot{B} = 0$. Therefore, $\ddot{B} = 0$, which is the consistency of $\dot{B} = 0$, is automatically satisfied by $\dot{H} = \dot{B} = 0$. (This is no surprising, since (\ref{EQB}--\ref{EQconst}) form a constrained system and thus any valid solution of \eqref{EQconst} and one of the remaining two should automatically satisfy the other.)
Imposing \eqref{dS-H} and \eqref{dS-B} simplifies \eqref{EQB}, \eqref{EQHdot} and \eqref{EQconst} to
\begin{eqnarray}
&& \bar G_2 - 3 \left[ \bar G_{2,Y} - 2 \left( \bar G_{2,W1} + 3 \, \bar G_{2,W2} + \bar G_{2,W3} \right) B^2 \right] H^2 B^2 + 6 \bar G_4 H^2 \approx 0
\label{EQBGdS1}\\
&& \bar G_{2,X} + 2 \left[ \bar G_{2,Y} - \left( \bar G_{2,W_1} + 3 \bar G_{2,W_2} + \bar G_{2,W_3} \right) B^2 \right] H^2 + 6 \bar G_{4,X} H^2 \approx 0
\label{EQBGdS2}
\end{eqnarray}
with the third equation being a redundant one, due to the nature of a constrained system.
Bearing in mind that $\bar G_{2}$ and $\bar G_4$ are functions of $B$ and $H$, \eqref{EQBGdS1} and \eqref{EQBGdS2} are simply algebraic equations of these variables, and the solutions to them provide the de Sitter background in the considered model.

Given a solution to this set of equations, (\ref{dS-H}--\ref{EQBGdS2}), we impose the conditions under which it is an attractor of the system. To this end, we perturb $B$ and $H$ as $B = B_0 + \delta B$ and $H = H_0 + \delta H$ in the original equations (\ref{EQB}--\ref{EQconst}) such that $B_0$ and $H_0$ satisfy (\ref{dS-H}--\ref{EQBGdS2}) and then linearize them in terms of $\delta B$ and $\delta H$. Now $\delta H$ (and $\delta\dot H$) can be algebraically solved in favor of $\delta B$, $\delta\dot B$ and $\delta\ddot B$, giving the master equation with respect to $\delta B$ with constant coefficients. This second-order equation takes the form $\delta\ddot B + 3 H_0 \delta\dot B + C_{\rm att} H_0^2 \delta B = 0$, where the numerator and denominator of $C_{\rm att}$ are given by
\begin{eqnarray}
\left( {\rm Num. \; of \;} C_{\rm att} \right) & = &
G_W \left[ 4 G_4 + 6 G_{4,X} B_0^2 + 9 G_{4,XX} B_0^4 + \frac{3 G_{2,XX}}{2 H_0^2} \, B_0^4 \right] B_0^2
\nonumber\\ &&
+ 3 G_{W,X} \left[ 2 \left( G_4 - 3 G_{4,X} B_0^2 \right) - 3 \left( G_{2,Y} + G_{2,XY} B_0^2 + G_{2,YY} H_0^2 B_0^2 \right) B_0^2 \right] B_0^4
\nonumber\\ &&
+ 3 G_{W,Y} \left[ 2 \left( 5 G_4 + 9 G_{4,X} B_0^2 + 9 G_{4,XX} B_0^4 \right) + \left( 2 G_{2,Y} + \frac{3 G_{2,XX}}{H_0^2} \, B_0^2 + 3 G_{2,XY} B_0^2 \right) B_0^2 \right] H_0^2 B_0^4
\nonumber\\ &&
- 3 \left( G_{W,W_1} + 3 G_{W,W_2} + G_{W,W_3} \right) \bigg[ 2 \left(4 G_4 + 12 G_{4,X} B_0^2 + 9 G_{4,XX} B_0^4 \right)
\nonumber\\ && \qquad\qquad\qquad\qquad\qquad\qquad\qquad\;\;\;
+ 5 G_{2,Y} B_0^2 + 3 \left( \frac{G_{2,XX}}{H_0^2} + 2 G_{2,XY} + G_{2,YY} H_0^2 \right) B_0^4 \bigg] H_0^2 B_0^6
\nonumber\\ &&
+ 6 G_W \left[ G_{W,X} + \left( G_{W,W_1} + 3 G_{W,W_2} + G_{W,W_3} \right) H_0^2 B_0^2 \right] B_0^8
+ 9 \left( G_{W,X} + G_{W,Y} H_0^2 \right)^2 B_0^{10}
\nonumber\\ &&
+ \bigg[ 3 G_4 \left( 6 G_{4,XX} + \frac{G_{2,XX}}{H_0^2} + G_{2,XY} - 2 G_{2,YY} H_0^2 \right)
\nonumber\\ && \quad\;\;
+ 3 G_{4,X} \left( G_{2,Y} + 3 G_{2,XY} B_0^2 - 3 G_{2,YY} H_0^2 B_0^2 \right) - \frac{9}{2} \, G_{4,XX} \left( G_{2,Y} + 3 G_{2,YY} H_0^2 B_0^2 \right) B_0^2
\nonumber\\ && \quad\;\;
+ G_{2,Y}^2 - \frac{3}{4} \, G_{2,Y} \left( \frac{G_{2,XX}}{H_0^2} - 4 G_{2,XY} \right) B_0^2 + \frac{9}{4} \, \left( G_{2,XY}^2 - G_{2,XX} G_{2,YY} \right) B_0^4 \bigg] B_0^2
\; ,
\nonumber\\
\left( {\rm Den. \; of \;} C_{\rm att} \right) & = &
- 4 G_{4,X}^2 B_0^2 +
\left( G_4 + 2 G_{4,X} B_0^2 \right) \Big\{ 12 \left[ G_{W,Y} - \left( G_{W,W_1} + 3 G_{W,W_2} + G_{W,W_3} \right) B_0^2 \right] H_0^2 B_0^4
\nonumber\\ && \qquad\qquad\qquad\qquad\qquad\qquad\qquad
+ 2 G_W B_0^2 - \left( G_{2,Y} + 3 G_{2,YY} H_0^2 B_0^2 \right) \Big\}\; ,
\label{cond-attractor}
\end{eqnarray}
with $G_W \equiv \bar G_{2,W_1} + 3 \bar G_{2,W_2} + \bar G_{2,W_3}$ and constant $B_0$ and $H_0$.
This parameter $C_{\rm att}$ plays the roll of the effective mass square (in the unit of $H_0$) in the conventional models, and in order to realize a sufficiently long period of inflation with a graceful exit, we require $\vert C_{\rm att} \vert \ll 1$. Whenever we consider the de-Sitter limit in the following analyses, we therefore take $C_{\rm att} \rightarrow 0$. In this limit, we have $\delta \ddot{B} + 3 H_0 \delta \dot{B} = 0$, leading to $\delta B \propto \delta \dot{B} \propto a^{-3}$ (with the constant solution absorbed into $B_0$), and thus deviations from de-Sitter solutions $B_0$ are quickly washed away. Since $\delta H$ and $\delta\dot H$ are written as linear combinations of $\delta B$, $\delta\dot B$ and $\delta\ddot B$, this guarantees an exponential decay of these variables as well and therefore the attractor behavior.%
\footnote{The procedure described here has another different but equivalent version that leads to the same attractor condition. We have solved the constraint equation in favor of $\delta B$ and its derivatives, but we can instead solve it in such a way that we have two coupled first-order equations in terms of $\delta H$ and $\partial_t (\delta B)$. The characteristic equation of this system is given as $\lambda^2 +3 \lambda + C_{\rm att} = 0$ with $\lambda$ being the eigenvalues and with the same $C_{\rm att}$ as given in \eqref{cond-attractor}. Some other versions are also possible.}
Hence we conclude that inflationary expansion is achieved by the approximate solutions of (\ref{dS-H}--\ref{EQBGdS2}) under the attractor condition $\vert C_{\rm att} \vert \ll 1$.
The deviation from the pure de Sitter is expected to be at slow-roll orders ${\cal O}(\dot H / H^2, \dot B / HB)$.
The condition for this attractor to stay isotropic, i.e.~$\langle A_i^{(a)} \rangle \propto \delta^a_i$ and $\langle g_{ij} \rangle \propto \delta_{ij}$, is shown in Sec.~\ref{subsec:isotropy}.

\section{Perturbations}
\label{sec:perturbations}

We have obtained in the previous section inflationary solutions given by (\ref{dS-H}--\ref{EQBGdS2}) in the theory \eqref{action} and their attractor condition $\vert C_{\rm att} \vert \ll 1$ in \eqref{cond-attractor}.
In this section, we analyze the linear perturbations around this background and search for the model regions where it is stable.
The concerned system has three vector fields and one metric, and thus the starting number of variables of perturbations is $3 \times 4 + 10 = 22$. A convenient decomposition for perturbations $\delta A^{(a)}_\mu \equiv A^{(a)}_\mu - \langle A^{(a)}_\mu \rangle$ and $\delta g_{\mu\nu} \equiv g_{\mu\nu} - \langle g_{\mu\nu} \rangle$ respecting the background configuration \eqref{A-vev} and \eqref{FLRW} is \cite{Maleknejad:2011jw}
\begin{eqnarray}
&& \delta A_0^{(a)} = Y_a + \partial_a Y \; , \quad
\delta A_i^{(a)} = a(t) \left[ \delta Q \, \delta_{ai} + \partial_i \left( M_a + \partial_a M \right) + \epsilon_{iab} \left( U_b + \partial_b U \right) + t_{ia} \right] \; ,
\label{decomp-vector}\\
&& \delta g_{00} = 2 \phi \; , \quad
\delta g_{0i} = a(t) \left( B_i + \partial_i B \right) \; , \quad
\delta g_{ij} = a^2(t) \left( 2 \psi \, \delta_{ij} + 2 \partial_i \partial_j E + \partial_i E_j + \partial_j E_i + h_{ij} \right)
\; ,
\label{decomp-metric}
\end{eqnarray}
where we reserve the indices $a,b,\dots$ for different vector fields, while $i,j,\dots$ for the spacial coordinates. The ``vector'' $(Y_a , M_a, U_a, B_i, E_i)$ and ``tensor'' $(t_{ia} , h_{ij})$ modes have transverse/traceless properties, i.e.
\begin{eqnarray}
&& \partial_i Y_i = \partial_i M_i = \partial_i U_i = \partial_i B_i = \partial_i E_i = 0 \; , \\
&& \partial_i t_{ij} = \partial_j t_{ij} = \partial_i h_{ij} = \partial_j h_{ij} = 0 \; , \quad t_{ii} = h_{ii} = 0 \; ,
\end{eqnarray}
and also $t_{ij}$ and $h_{ij}$ are symmetric. We denote other modes as ``scalar.''
While the modes $\{Y, \delta Q, M, U \}$, $\{ Y_a, M_a, U_a \}$ and $\{ t_{ia} \}$ do not transform as scalar, vector and tensor, respectively, in the standard sense, the decomposition \eqref{decomp-vector} is particularly useful in that, already at the linear order, they are decoupled from each other and coupled separately to the true scalar $\{\phi , B , \psi, E\}$, vector $\{B_i, E_i\}$ and tensor $\{h_{ij}\}$ modes in the metric.
This is therefore an exceptional case to the standard decomposition theorem in that the decomposition is done with respect not only to the coordinate space but also to the vector-fields' internal space, which is also the case for models like Chromo-natural inflation and Gauge-flation.
The quadratic action can be written as a sum of the three sectors,
\begin{equation}
S^{(2)} = S^{(2)}_S \left[ Y, \delta Q, M, U, \phi, B, \psi, E \right] + S^{(2)}_V \left[ Y_a , M_a, U_a, B_i, E_i \right] + S^{(2)}_T \left[ t_{ia} , h_{ij} \right] \; .
\label{quadaction-before}
\end{equation}
Let us note in passing that the decomposition of the type (\ref{decomp-vector}--\ref{decomp-metric}) can always be done, but its usefulness comes from this linear decoupling, a consequence of the background field configuration \eqref{A-vev} and FLRW spacetime (\ref{FLRW}).

Not all of the modes are physically relevant degrees of freedom. There is a gauge freedom regarding general coordinate transformation, which we fix by taking the spatially-flat gauge, $\psi = E = E_i = 0$. Note that the vector fields under consideration do not possess gauge invariance and no further gauge fixing is applicable.
Among the remaining modes, $\{Y,\phi,B\}$ and $\{Y_a, B_i\}$ are non-dynamical, i.e.,~they enter the quadratic action without time derivatives, up to total derivatives, and therefore can be integrated out by constraint equations.
We are then left with the truly propagating degrees of freedom, $3$ scalar $\{\delta Q , M , U \}$, $4$ vector $\{ M_a, U_a \}$ and $4$ tensor $\{t_{ia} , h_{ij} \}$ modes.
The action \eqref{quadaction-before} reduces to
\begin{equation}
S^{(2)} = S^{(2) \, \prime}_S \left[ \delta Q, M, U \right] + S^{(2) \, \prime}_V \left[ M_a, U_a \right] + S^{(2)}_T \left[ t_{ia} , h_{ij} \right] \; ,
\label{quadaction-after}
\end{equation}
expressed in terms only of physical degrees of freedom.
In the current work, we are only interested in studying the linearized theory of perturbations. This amounts to expanding the equations of motion up to first order, or equivalently to expanding the action up to quadratic order. In either way, the three sectors (scalar/vector/tensor) are decoupled from each other (their couplings emerge only at the nonlinear level).
We study the stability conditions in each sector; we first outline the general procedure and then proceed the analyses of each sector separately in the following subsections.
Also we explain that the stability of the tensor modes in the long-wavelength limit is equivalent to the one against anisotropic deformations in our setup and demonstrate to ensure the background isotropy by demanding the stability of the tensor sector in the this limit $k \rightarrow 0$.

\subsection{General procedure}
\label{subsec:general}

Our goal is to investigate the stability of the inflationary solution (\ref{dS-H}--\ref{EQBGdS2}) against perturbations. Disastrous breakdown of the solution arises in the form of ghost and/or gradient instabilities. We focus in particular on their high momentum limit, as this would immediately lead to the breakdown of vacuum states while low-energy instabilities only hint the ones {\it \`{a} la} Jeans instability and may thus be under control \cite{Gumrukcuoglu:2016jbh}.
This subsection is devoted to outlining the procedure of how to identify those ghost and gradient instabilities and the conditions to evade them.

In performing the computations, it is convenient to Fourier-transform the perturbations as
\begin{equation}
\delta \left( t , \bm{x} \right) = \int \frac{d^3 k}{\left( 2 \pi \right)^{3/2} } \, {\rm e}^{i \bm{k} \cdot \bm{x}} \, \delta \left( t , \bm{k} \right) \; ,
\label{FT}
\end{equation}
where $\delta \left( t, \bm{x} \right)$ denotes each perturbation. The variations with respect to non-dynamical modes ($\{Y, \phi, B\}$ in the scalar sector and $\{Y_a, B_i \}$ in the vector) give the constraint equations, which are algebraic in the Fourier space, and they can be solved to integrate out those modes.
As a result, the action of each decoupled sector (sometimes sub-sector) is written in terms only of dynamical modes, i.e.~$\{ \delta Q , M ,U \}$ for scalar, $\{ M_a, U_a \}$ for vector and $\{ t_{ia} , h_{ij} \}$ for tensor, and can be cast into the form
\begin{equation}
S^{(2)} = \frac{1}{2} \int dt \, d^3k \, a^3 \left[ \dot\delta^\dagger_D \, T \, \dot\delta_D + \dot\delta^\dagger_D \, X \, \delta_D - \delta^\dagger_D \, X \, \dot\delta_D - \delta_D^\dagger \, \Omega^2 \, \delta_D \right] \; ,
\label{quadact-dynam}
\end{equation}
up to addition of appropriate total derivatives, where $\delta_D$ is an array of dynamical modes, dagger denotes Hermitian conjugate, and $T$, $X$ and $\Omega^2$ are square matrices with the size corresponding to the number of the dynamical modes and with the properties $T^\dagger = T$, $X^\dagger = - X$ and $\Omega^{2 \, \dagger} = \Omega^2$. The non-vanishing off-diagonal components of these matrices are the indication of a coupled system. Notice that we factor out the volume element ($a^3$) from the definitions of the matrices.

The matrix $T$ is in general not diagonal. We thus perform a rotation in the field space
\begin{equation}
\delta_D = R \, \Delta_D \; ,
\label{rotation}
\end{equation}
such that $R^\dagger T R$ is diagonal.
Then the action \eqref{quadact-dynam} becomes, again after properly adding total derivatives,
\begin{equation}
S^{(2)} = \frac{1}{2} \int dt \, d^3k \, a^3 \left[ \dot\Delta^\dagger_D \, \bar T \, \dot\Delta_D + \dot\Delta^\dagger_D \, \bar X \, \Delta_D - \Delta^\dagger_D \, \bar X \, \dot\Delta_D - \Delta^\dagger_D \, \bar \Omega^2 \, \Delta_D \right] \; ,
\label{quadact-diag}
\end{equation}
where $\bar X^\dagger = - \bar X$, $\bar\Omega^{2\,\dagger} = \bar\Omega^2$, and
\begin{equation}
\bar T \equiv R^\dagger T R = {\rm diag} \left( \lambda_1 , \lambda_2 , \dots , \lambda_N \right) \; ,
\end{equation}
with $N$ the number of dynamical degrees.
The matrix $\bar T$ is Hermitian by construction, and thus $\lambda_i$ are real for all $i$.
It is convenient for our later purpose to choose the rotation matrix \eqref{rotation} so to have (we have this freedom for constant rescaling)
\begin{equation}
\lambda_i \propto k^0 \; , \qquad \mbox{for } \; k \rightarrow \infty \; .
\label{lambda-highk}
\end{equation}
Ghosts refer to the modes that have a negative time kinetic term, and therefore, the no-ghost condition is the requirement
\begin{equation}
\lambda_i > 0 \; , \qquad
\mbox{no ghost condition} \; ,
\label{noghost}
\end{equation}
for all $i$.
As mentioned at the beginning of this subsection, we are particularly interested in the cases where \eqref{noghost} is satisfied in large $k$ limit.

The equations of motion for $\Delta_D$ are derived by varying \eqref{quadact-diag} with respect to $\Delta^\dagger_D$.
Thanks to the choice \eqref{lambda-highk} and to the fact that $\bar X \propto k^0 \, {\rm or} \, k^1$ and $\Omega^2 \propto k^2$ in the $k \rightarrow \infty$ limit for all the cases of our current concern, the equations of motion in this limit read
\begin{equation}
\left( \bar T \, \partial_t^2 + 2 \bar X \, \partial_t + \bar \Omega^2 \right) \Delta_D \approx 0 \; ,
\label{EOM-subhorizon}
\end{equation}
where time derivatives of $\bar T$ and $\bar X$ are of order $\dot H$ or $\dot B$ and therefore negligible. Eq.~\eqref{EOM-subhorizon} is an operator equation, and we assume that the ``eigenvector'' takes the form $\Delta_D \propto \exp \left[ - i \int^t c_s k / a(t') \, dt' \right] \bm{e}$ in the deep subhorizon, where $c_s$ is the sound speed and $\bm{e}$ is a constant vector.
Then the existence of non-trivial solutions to \eqref{EOM-subhorizon} requires
\begin{equation}
\det\left[ \bar T \, \frac{k^2}{a^2} \, c_s^2 + 2 i \bar X \, \frac{k}{a} \, c_s - \bar \Omega^2 \right] = 0 \; ,
\label{cs_det}
\end{equation}
where the time derivatives of $c_s$ and $a$ are of order $H$, $\dot H$ or $\dot B$, and therefore negligible compared to $k$ in this limit.
Solving \eqref{cs_det} determines the sound speed of the dynamical modes in large $k$ limit, and the condition to evade gradient instabilities is given by $c_s^2 > 0$ for all the solutions of \eqref{cs_det}.
For our current study of perturbations, each decoupled sector is either a $2\times2$ or $3\times3$ system. For a $2\times2$ case, \eqref{cs_det} leads to a quadratic equation for $c_s^2$,
\begin{equation}
\left( c_s^2 \right)^2 - \alpha \, c_s^2 + \beta = 0 \; .
\end{equation}
To ensure that the two roots of $c_s^2$ for this equation are both real and positive, we thus require
\begin{equation}
\alpha > 0 \; , \quad
\beta > 0 \; , \quad
\alpha^2 - 4 \beta \ge 0 \; , \qquad
\mbox{for $c_s^2 > 0$ in a $2\times2$ system} \; .
\label{gradstab-2by2}
\end{equation}
On the other hand, for a $3\times3$ system, \eqref{cs_det} reduces to a cubic equation of $c_s^2$,
\begin{equation}
\left( c_s^2 \right)^3 - \gamma \left( c_s^2 \right)^2 + \delta \, c_s^2 - \epsilon = 0
\label{cs2-char3}
\end{equation}
and then the reality and positivity of all the roots of $c_s^2$ imposes
\begin{equation}
\gamma > 0 \; , \quad  \delta > 0 \; , \quad \epsilon > 0 \; , \quad \gamma^2 - 3 \, \delta \ge 0 \; , \quad
\gamma^2 \delta^2 - 4 \delta^3 - 4 \gamma^3 \epsilon + 18 \gamma \delta \epsilon - 27 \epsilon^2 \ge 0 \; , \qquad
\mbox{for $c_s^2 > 0$ in a $3\times3$ system} \; ,
\label{gradstab-3by3}
\end{equation}
In order for the system to be completely stable against ghost and gradient instabilities, all the dynamical degrees of freedom must simultaneously satisfy both of the conditions of no-ghost \eqref{noghost} and positive squared sound speeds \eqref{gradstab-2by2}/\eqref{gradstab-3by3}.
In the following subsections, we apply this procedure for the tensor, vector and scalar sectors individually. The conditions for positive sound speeds are extremely lengthy for the vector and scalar sectors with the general functions of $G_2$ and $G_4$ (see \eqref{action}), and so we restrict ourselves to showing the explicit results only in Sec.~\ref{sec:examples}, where we demonstrate some illustrative examples.

\subsection{Tensor sector}
\label{subsec:tensor}

The tensor sector perturbations consist of $h_{ij}$ and $t_{ia}$, with $2+2=4$ degrees of freedom.
While $t_{ia}$ is not a tensor quantity under a spatial rotation, it couples to the metric tensor mode $h_{ij}$ at the linear level.
Due to the presence of $Z = - F_{\mu\nu}^{(a)} \tilde F^{(a), \mu\nu} / 4$ in the function $G_2$ (see \eqref{def-XYZW}), parity violation is involved in this theory. Thus it is convenient to work in the right- and left-handed basis of the tensor modes, following the decomposition of $h_{ij}$ as
\footnote{The usual decomposition of tenor perturbations $h_{ij}=h_{+}e_{ij}^{+}+h_{\times} e_{ij}^{\times}$ is translated in terms of the left-handed and right-handed canonical modes by $h_R=\left(h_+-ih_{\times}\right)/2$, $h_L=\left(h_++ih_{\times}\right)/2$.}
\begin{equation}
h_{ij} \left( t , \bm{k} \right) = \sum_{\lambda = R,L} \Pi_{ij}^{\lambda} \left( \hat k \right) h_\lambda \left( t , \bm{k} \right) \; , \quad
t_{ia} \left( t , \bm{k} \right) = \sum_{\lambda = R,L} \Pi_{ia}^{\lambda} \left( \hat k \right) t_\lambda \left( t , \bm{k} \right) \; ,
\end{equation}
where $\Pi_{ij}^\lambda$ is the polarization tensor satisfying symmetric, traceless and transverse properties together with $\Pi_{ij}^{R/L} ( - \hat k ) = \Pi_{ij}^{R/L \, *} ( \hat k ) = \Pi_{ij}^{L/R} ( \hat k )$ and $i \epsilon_{ikl} \hat k_k \Pi_{lj}^{R/L} (\hat k) = \pm \Pi_{ij}^{R/L} (\hat k)$.
After this decomposition, we observe that the two handed modes are separated, namely
\begin{equation}
S^{(2)}_T \left[ h_{ij} , t_{ia} \right] = S^{(2)}_R \left[ h_R , t_R \right] + S^{(2)}_L \left[ h_L , t_L \right] \; .
\label{action-tensor}
\end{equation}
The actions of the two sectors do not exactly coincide, which is indication of parity violation, and the difference appears only in the terms that are proportional to $\bar G_{2,Z}$.
However, they do not contribute in the high $k$ limit and do not arise in either the no-ghost or gradient stability conditions for tensor modes. Therefore this difference between the right- and left-handed tensor modes is irrelevant for our current analysis, and we hereafter omit the labels $R/L$ whenever causing no ambiguity.

In $S^{(2)}_{R/L}$, the modes $h_{R/L}$ and $t_{R/L}$ have kinetic mixing. Following the procedure described in Sec.~\ref{subsec:general}, we diagonalize the kinetic matrix by the rotation
\begin{equation}
R_T = \left(
\begin{array}{cc}
\bar G_{2,Y} - 2 \left( \bar G_{2,W_1} + \bar G_{2,W_3} \right) B^2 & 0 \\
\bar G_{4,X} B & \frac{1}{2}
\end{array}
\right) \; , \label{eqn:tensor-mixing}
\end{equation}
where subscript $T$ denotes tensor modes.
Then the kinetic matrix after the rotation, corresponding to $\bar T$ in \eqref{quadact-diag}, becomes diagonal. Its explicit form is
\begin{equation}
\bar T_T = \left(
\begin{array}{cc}
2 \, C_{\rm NG1} C_{\rm NG2} & 0 \\
0 & C_{\rm NG1}
\end{array}
\right) \; ,
\label{Tmat-T}
\end{equation}
where
\begin{equation}
C_{\rm NG1} \equiv \bar G_{2, Y} - 2 \left( \bar G_{2,W_1} + \bar G_{2,W_3} \right) B^2 \; , \quad
C_{\rm NG2} \equiv C_{\rm NG1} \left( \bar G_4 + 2 \, \bar G_{4,X} B^2 \right) - 2 \, \bar G_{4,X}^2 B^2 \; .
\label{CNG12}
\end{equation}
Notice that $\bar T_T$ is independent of $k$.
In large $k$ limit, the mixing matrix $\bar X_T$ and mass matrix $\Omega^2_T$ read
\begin{equation}
\bar X_T \simeq 0 \; , \quad
\bar \Omega^2_T \simeq \frac{k^2}{a^2} \left(
\begin{array}{cc}
2 C_{\rm NG1}
\left[ C_{\rm NG1} \left( G_4 + 3 G_{4,X} B^2 \right) - 2 G_{4,X}^2 B^2 \right] -8 G_{2,W_1} G_{4,X}^2 B^4
& - 4 \bar G_{2,W_1} \bar G_{4,X} B^3 \\
- 4 \bar G_{2,W_1} \bar G_{4,X} B^3
& C_{\rm NG1} - 2 G_{2,W_1} B^2
\end{array}
\right) \; .
\label{X-O2-T-highk}
\end{equation}
Then the no-ghost conditions are given by requiring every component of $\bar T_T$ to be positive, i.e.,
\begin{equation}
C_{\rm NG1} > 0 \; , \quad
C_{\rm NG2} > 0 \; ,
\label{noG-tensor}
\end{equation}
and the no-gradient-instability conditions are obtained by demanding the positive roots of \eqref{cs_det} for $c_s^2$ given the large $k$ limit of the matrices, \eqref{Tmat-T} and \eqref{X-O2-T-highk}, yielding
\begin{eqnarray}
&& \alpha_T \equiv 2 + \frac{C_{\rm NG1} G_{4,X} - 2 G_{2,W_1} \left( G_4 + 2 G_{4,X} B^2 \right)}{C_{\rm NG2}} \, B^2 > 0
\label{gradstab-T1}\\
&& \beta_T \equiv 1 +  \frac{C_{\rm NG1} G_{4,X} - 2 G_{2,W_1} \left( G_4 + 3 G_{4,X} B^2 \right)}{C_{\rm NG2}} \, B^2 > 0 \; .
\label{gradstab-T2}\\
&& \alpha_T^2 - 4 \beta_T = \frac{B^4}{C_{\rm NG2}^2}
\left[
C_{\rm NG1}^2 G_{4,X}^2
+ 4 G_{2,W_1}^2 \left( G_4 + 2 G_{4,X} B^2 \right)^2
+ 4 G_{2,W_1} G_{4,X} \left( C_{\rm NG2} - 2 G_{4,X}^2 B^2 \right)
\right] > 0 \; .
\label{gradstab-T3}
\end{eqnarray}
As mentioned previously, these conditions apply to both of the right- and left-handed modes, and therefore the stability in the tensor sector is guaranteed by satisfying all of (\ref{noG-tensor}--\ref{gradstab-T3}) simultaneously.

In the above, we have demonstrated the calculation of the tensor stability and shown the explicit expressions for all the conditions, as they are relatively compact in length. We now proceed to the vector and scalar sectors and follow the same procedures to obtain their stability conditions, but the expressions for the sound speeds for them are extremely lengthy and not illuminating. Thus we refrain from showing the full results for the most general functions of $G_2$ and $G_4$. Instead, we only show the no-ghost conditions in the following subsections and discuss the gradient stability conditions when we study some specific examples in Sec.~\ref{sec:examples}.

\subsection{Vector sector}
\label{subsec:vector}

The vector sector is initially composed of $2 \times 4 = 8$ perturbations, $Y_a,M_a,U_a,B_i$ ($E_i$ have been gauged away), including the non-dynamical modes $Y_a$ and $B_i$. As we have used the $R/L$ basis in the tensor sector, it is convenient to work the circular polarization states. This amounts to decomposing the vector modes $V_i = \{ Y_i,M_i,U_i,B_i \}$ as, in the Fourier space,\footnote{More explicitly, provided that momentum vector $V_i$ is oriented along the $z$-axis, the components $V_{\pm}$ can be expressed in terms of $V_i$ as $V_\pm=\left(V_1\mp iV_2\right)/2$.}
\begin{equation}
V_i \left( t , \bm{k} \right) = \sum_{\lambda=\pm} e^\lambda_i \left( \hat k \right) V_{\lambda} \left( t , \bm{k} \right) \; ,
\end{equation}
where $e^\lambda_{i}$ is the circular polarization vector satisfying traceless property and $e^\pm_i ( - \hat k ) = e^{\pm \, *}_i ( \hat k ) = e^\mp_i ( \hat k )$ and $i \epsilon_{ijk} \hat k_j e^\pm_k ( \hat k ) = \pm e^\pm_i ( \hat k )$.
Then the $\pm$ sectors decouple, i.e.,
\begin{equation}
S^{(2)}_V \left[ Y_a , M_a, U_a, B_i \right] = S^{(2)}_+ \left[ Y_+ , M_+, U_+, B_+ \right] + S^{(2)}_- \left[ Y_- , M_-, U_-, B_- \right] \; .
\end{equation}
One can observe that the two sub-sectors $S^{(2)}_\pm$ do not coincide. This is due to the fact that the non-vanishing $Z$ dependence in $G_2$ breaks parity (the same reason as $S^{(2)}_R \ne S^{(2)}_L$ in the tensor) and also that the $U_a$ mode behaves as a pseudo-vector due to the presence of $\epsilon_{iab}$ in front of its definition, \eqref{decomp-vector}, with the background configuration \eqref{A-vev}. Therefore, the difference between $S^{(2)}_+$ and $S^{(2)}_-$, hinting parity violation, comes with the terms either proportional to $\bar G_{2,Z}$ or linear in $U_\pm$.

To proceed to stability analyses, we first integrate out the non-dynamical modes, $Y_\pm$ and $B_\pm$. They enter the quadratic action without time derivatives, up to addition of total derivatives, and thus the variations of action with respect to them lead to constraint equations that are only algebraic in the Fourier space. Solving them and plug the expressions for $Y_\pm$ and $B_\pm$ back in, we obtain the action only in terms of the physical degrees of freedom, i.e.,
\begin{equation}
S^{(2)}_\pm \left[ Y_\pm , M_\pm, U_\pm, B_\pm \right] = S^{(2) \, \prime}_\pm \left[ M_\pm, U_\pm \right] \; ,
\end{equation}
where the equality holds on the constrained hypersurface.
Since $M_\pm$ and $U_\pm$ have kinetic mixing, we follow the procedure in Sec.~\ref{subsec:general} and rotate the system by
\begin{equation}
R_\pm = \left(
\begin{array}{cc}
\frac{1}{2 \, a} & \mp \frac{1}{k} \left[ \bar G_{2, Y} - 2 \left( \bar G_{2,W_1} - \bar G_{2,W_3} \right) B^2 \right] \left( \bar G_4 + \bar G_{4,X} B^2 \right) \\
0 & \left( \bar G_{2,Y} - 2 \bar G_{2,W_1} B^2 \right) \left( \bar G_4 + 2 \bar G_{4,X} B^2 \right) - \bar G_{4,X}^2 B^2
\end{array}
\right)
\end{equation}
where subscripts $\pm$ denote the $\pm$ vector modes.
Then the kinetic matrices after the rotation, corresponding to $\bar T$ in \eqref{quadact-dynam}, become
\begin{equation}
\bar T_\pm =
\left(
\begin{array}{cc}
\frac{H^2 C_{\rm NG3}}{
1 +
2 \, \frac{a^2}{k^2} \, C_{\rm NG3} C_{\rm NG4}^{-1}
\left( G_4 + G_{2,Y} B^2 - 2 G_{2,W_1} B^4 \right)
} & 0 \\
0 & 2 C_{\rm NG2} C_{\rm NG4} C_{\rm NG5}
\end{array}
\right)
\end{equation}
where
\begin{eqnarray}
C_{\rm NG3} & \equiv &
\left[ - \bar G_{2,Y} + 2 \left( 2 \bar G_{2,W_1} + 2 \bar G_{2,W_2} + \bar G_{2,W_3} \right) B^2 \right] \left( 1 + \frac{\dot B}{HB} \right)^2
\nonumber\\
&& \;\;
- 2 \left( \bar G_4 + \bar G_{4,X} B^2 \right) \frac{\dot H}{H^2 B^2}
+ 6 \bar G_{4,XX} \frac{B \dot B}{H}
+ \frac{2}{H^2 B^2} \, \partial_t \left( \bar G_{4,X} B \dot B \right) \; ,
\label{CNG3}\\
C_{\rm NG4} & \equiv & \left( \bar G_{2,Y} - 2 \bar G_{2,W_1} B^2 \right) \left( \bar G_4 + 2 \bar G_{4,X} B^2 \right) - \bar G_{4,X}^2 B^2 \; ,
\label{CNG4}\\
C_{\rm NG5} & \equiv & \bar G_{2, Y} - 2 \left( \bar G_{2,W_1} - \bar G_{2, W_3} \right) B^2 \; .
\label{CNG5}
\end{eqnarray}
To obtain \eqref{CNG3}, eq.~\eqref{EQHdot} is used to replace $\bar G_{2,X}$.
Notice that the kinetic matrices of both sub-sectors are identical, $\bar T_+ = \bar T_-$.
The no-ghost conditions in the vector sector are to ensure that both of the components of $\bar T_\pm$ should be strictly positive at all times. As argued at the beginning of Sec.~\ref{subsec:general}, we pay particular attention in the large momentum limit, $k \rightarrow \infty$. Therefore, the vector ghost stability conditions are
\begin{equation}
C_{\rm NG3} > 0 \; , \quad
C_{\rm NG4} C_{\rm NG5} > 0 \; ,
\label{noG-vector}
\end{equation}
provided $C_{\rm NG2} > 0$, already imposed from the tensor no-ghost condition \eqref{noG-tensor}.

In large $k$ limit, one can observe that $\bar T_+ = \bar T_-$, $\bar X_+ = - \bar X_-$ and $\bar\Omega^2_+ = \bar\Omega^2_-$, and for this reason, the characteristic equation \eqref{cs_det} for sound speed is identical for the $\pm$ modes. Therefore, the conditions to ensure gradient stability are the same for the two sub-sectors and can be obtained by \eqref{gradstab-2by2} with corresponding expressions of $\alpha$ and $\beta$, denoting $\alpha_V$ and $\beta_V$, respectively. We discuss these conditions in more details for some specific cases in Sec.~\ref{sec:examples}.

\subsection{Scalar sector}
\label{subsec:scalar}

The scalar sector contains $6$ perturbations, $Y, \delta Q, M, U, \phi$ and $B$, with $\psi$ and $E$ gauged away. Among these modes, $Y$, $\phi$, and $B$ are non-dynamical and enter the quadratic action without any time derivatives (up to total derivatives). By solving the constraint equations, obtained by varying the action $S^{(2)}_S$ with respect to them, we can integrate out these non-dynamical modes and express $S^{(2)}_S$ in terms only of the dynamical variables, $\{\delta Q, M, U\}$, i.e.
\begin{equation}
S^{(2)}_S \left[ Y, \delta Q, M, U, \phi, B \right] = S^{(2) \, \prime}_S \left[ \delta Q, M, U \right] \; ,
\end{equation}
where the equality holds on the constrained hypersurface.
One can then observe that $U$ does not have a kinetic mixing with $\delta Q$ or $M$, and so it suffices to rotate the $\left\{ \delta Q , M \right\}$ part. Together with a trivial rescaling for $U$, we rotate the system by, as in \eqref{rotation},
\begin{equation}
R_S = \left(
\begin{array}{ccc}
G_4 + G_{4,X} B^2 & 0 & 0 \\
\frac{C_{\rm NG6}}{2 k^2} \frac{2 G_4 + C_{\rm NG1} B^2}{G_4 + G_{4,X} B^2} & - \frac{1}{k a} \\
0 & 0 & \frac{1}{k}
\end{array}
\right)
\end{equation}
where subscript $S$ denotes the scalar sector,
\begin{equation}
C_{\rm NG6} \equiv \frac{\left\{ \bar G_{2,Y} - 2 \bar G_W B^2 + 3 \bar G_{2,YY} H^2 B^2 + 12 H^2 B^4 \left[ - \partial_Y + B^2 \left( \partial_{W_1} + 3 \partial_{W_2} + \partial_{W_3} \right) \right] \bar G_W \right\} \left( \bar G_4 + 2 \bar G_{4,X} B^2 \right) + 4 \bar G_{4,X}^2 B^2}{\bar G_{2,Y} - 2 \left( \bar G_{2,W_1} + \bar G_{2,W_2} + \bar G_{2,W_3} \right) B^2 + \bar G_{2,YY} H^2 B^2 + 4 H^2 B^4 \left[ - \partial_Y + B^2 \left( \partial_{W_1} + 3 \partial_{W_2} + \partial_{W_3} \right) \right] \bar G_W } \; ,
\label{CNG6}
\end{equation}
with $\bar G_W \equiv \bar G_{2,W_1} + 3 \bar G_{2,W_2} + \bar G_{2,W_3}$, and the operator $- \partial_Y + B^2 \left( \partial_{W_1} + 3 \partial_{W_2} + \partial_{W_3} \right)$ is acting only on $\bar G_W$ on its right.
Note that this expression already assumes the de Sitter solution of the background, see (\ref{dS-H}--\ref{EQBGdS2}).
Then the kinetic matrix after the rotation, corresponding to $\bar T$ in \eqref{quadact-dynam}, takes the form, in de Sitter and large $k$ limit,
\begin{equation}
\bar T_S = 2 \left(
\begin{array}{ccc}
C_{\rm NG2} C_{\rm NG6} & 0 & 0 \\
0 & H^2 C_{\rm NG7} + {\cal O} (k^{-2}) & 0 \\
0 & 0 & C_{\rm NG5}
\end{array}
\right) \; ,
\end{equation}
where
\begin{equation}
C_{\rm NG7} \equiv - \bar G_{2,Y} + 2 \left( 2 \bar G_{2,W_1} + 2 \bar G_{2, W_2} + \bar G_{2,W_3} \right) B^2 \; .
\label{CNG7}
\end{equation}
Notice $C_{\rm NG7} = C_{\rm NG3}$ in the de Sitter solutions we are currently interested in, $\dot B = \dot H = 0$, and therefore as far as the stability around such background is concerned, $C_{\rm NG7}$ does not give any new condition.
The stability against ghosts in the scalar sector then requires that the remaining two components in $\bar T_S$ be positive at all times, namely
\begin{equation}
C_{\rm NG5} > 0 \; , \quad C_{\rm NG6} > 0 \; ,
\label{noG-scalar}
\end{equation}
provided that $C_{\rm NG2} > 0$, which is ensured by the tensor stability \eqref{noG-tensor}.

In order to compute the stability against gradient terms, we take the $k \rightarrow \infty$ limit on the de Sitter background and then compute the characteristic equation \eqref{cs2-char3} for $c_s^2$. The system is now $3\times 3$, and therefore we impose $5$ conditions as in \eqref{gradstab-3by3}, for reality and positivity of all the $c_s^2$ values. We denote the three coefficients in \eqref{cs2-char3} by $\gamma_S$, $\delta_S$ and $\epsilon_S$. We show some concrete examples and illustrate these stability conditions explicitly in Sec.~\ref{sec:examples}.

\subsection{Attractor condition against anisotropic expansion}
\label{subsec:isotropy}

In this subsection, we derive the condition under which the isotropic configuration of the background, given by \eqref{A-vev} and \eqref{FLRW}, is stable against anisotropic deformations. As opposed to the case of a scalar inflaton, even if the system starts out with the configuration \eqref{A-vev}, off-diagonal components of $\langle A_i^a \rangle$ might be excited dynamically and spoil the isotropic expansion. The anisotropic part in the metric corresponds to its traceless and transverse components in the homogeneous background and therefore coincide exactly with $h_{ij}$ in \eqref{decomp-metric} in the long-wavelength limit. This tensor mode couples to $t_{ia}$ in $\delta A_i^a$ at the linear order. Therefore, in order to have the isotropic configuration, it suffices to guarantee the stability of the tensor sector $\{h_{ij} , t_{ia} \}$ in the limit $k \rightarrow 0$.%
\footnote{One may speculate that due to the specific form of the decomposition \eqref{decomp-vector}, the ``vector'' mode $U_a$ also enters in $\delta A_i^a$ without any spatial derivatives and thus could source the metric vector mode $B_i$, spoiling the isotropy. However, since $B_i$ is a vector while $U_a$ behaves as a pseudo-vector on the vev \eqref{A-vev}, they do not couple in the limit $k \rightarrow 0$ at the linear order, which we have confirmed explicitly. Therefore the vector sector would not induce any background anisotropy.}
It is worth emphasizing that our approach is slightly different from a conventional analysis of attractor behaviors in the context of the inflation which shows the FLRW attractor solution in the system starting from an anisotropic metric, e.g.~Bianchi type-I cosmology. Our method is, instead of adding an additional dynamical variable responsible for anisotropy at the background level, simply to consider the stability of the long-wavelength tensor modes. This should be satisfactory at least for small, linearized anisotropic deformations.

The tensor sector consists of two sub-sectors, right-handed $\{ h_R , t_R\}$ and left-handed $\{h_L , t_L \}$, and their actions are identical in the $k \rightarrow 0$ limit.
By varying the tensor action \eqref{action-tensor} with respect to $h_{R/L}$ and $t_{R/L}$, we obtain a set of two second-order differential equations in each sector. After differentiating them once and twice with respect to time, we can substitute all the $t_{R/L}$ and its derivatives to obtain a single fourth-order differential equation for $h_{R/L}$, which reads, in the de-Sitter and $k \rightarrow 0$ limit,
\begin{equation}
\partial_t^4 h_{R/L} + 6 H \, \partial_t^3 h_{R/L} + H^2 C_{h,2} \, \partial_t^2 h_{R/L} + H^3 C_{h,1} \, \partial_t h_{R/L} = 0 \; ,
\label{eq-h-4th}
\end{equation}
where
\begin{equation}
\begin{aligned}
C_{h,1} = & \frac{12 B^2}{C_{\rm NG2}} \bigg[
3 \left( \bar G_{2,W_1} + 2 \bar G_{2,W_2} + \bar G_{2,W_3} \right) \left( \bar G_{2,W_1} + 3 \bar G_{2,W_2} + \bar G_{2,W_3} \right) B^4
- 3 \left( 2 \bar G_{2,W_1} + 3 \bar G_{2,W_2} + 2 \bar G_{2,W_3} \right) \bar G_{4,X} B^2 \\ & \qquad\quad
- \left( \bar G_{2,W_1} + \bar G_{2,W_3} \right) \bar G_4
- \frac{1}{4} \, \bar G_{2,Y}^2
+ \frac{3}{2} \left( \bar G_{4,X} - \bar G_{2, W_2} B^2 \right) \bar G_{2,Y}
\bigg] \; , \\
C_{h,2} = & \frac{C_{h,1}}{3} + 9 \; .
\end{aligned}
\label{Ch}
\end{equation}
Let us note in passing that the absence of the term proportional to $h_{R/L}$ without time derivatives in \eqref{eq-h-4th} implies that it admits a constant solution for $h_{R/L}$ in the super horizon as in the standard slow-roll inflation in general relativity.
Constant tensor modes in super horizon can be regarded as gauge modes for local observers inside horizon and do not disturb the isotropic attractor, since the background perturbed by them become more and more indistinguishable from the unperturbed one as the universe expands at an accelerated rate.
The stability against anisotropic expansion is ensured by imposing that $h_{R/L}$ has only constant or decreasing modes in this limit. 
By taking the ansatz $h_{R/L} \propto \exp(\lambda H t)$, \eqref{eq-h-4th} reduces to a fourth-order polynomial equation, which can be solved as $\lambda = 0 , - 3, ( -3 \pm \sqrt{9 - 4 C_{h,1} / 3 } )/2$, using the relation \eqref{Ch} between $C_{h,1}$ and $C_{h,2}$. Therefore, the necessary and sufficient condition for $h_{R/L}$ to have only non-increasing modes, i.e.~for the real parts of all the $\lambda$ to be non-positive, imposes $C_{h,1} \ge 0$.%
\footnote{When $C_{h,1} = 27/4$, then two roots of $\lambda$ are degenerate. One can show that the solution to \eqref{eq-h-4th} in this case is $h_{R/L} \propto {\rm const.}, \, {\rm e}^{-3Ht},\, {\rm e}^{-3Ht/2},\, t\, {\rm e}^{-3Ht/2}$, and therefore it would not lead to any appreciable growth.} 
Therefore, in order to have isotropic expansion as an attractor against anisotropy, we require a single condition
\begin{equation}
\begin{aligned}
C_{\rm iso} &\equiv
3 \left( \bar G_{2,W_1} + 2 \bar G_{2,W_2} + \bar G_{2,W_3} \right) \left( \bar G_{2,W_1} + 3 \bar G_{2,W_2} + \bar G_{2,W_3} \right) B^4
- 3 \left( 2 \bar G_{2,W_1} + 3 \bar G_{2,W_2} + 2 \bar G_{2,W_3} \right) \bar G_{4,X} B^2 \\ & \quad\;
- \left( \bar G_{2,W_1} + \bar G_{2,W_3} \right) \bar G_4
- \frac{1}{4} \, \bar G_{2,Y}^2
+ \frac{3}{2} \left( \bar G_{4,X} - \bar G_{2, W_2} B^2 \right) \bar G_{2,Y} \\ &
\ge 0 \; ,
\end{aligned}
\label{Ciso}
\end{equation}
provided $C_{\rm NG2} > 0$.
This prevents the background system from running away from the isotropic configuration \eqref{A-vev}, and the expansion stays isotropic throughout the inflationary evolution.

\subsection{Stability conditions}
\label{subsec:stability-all}

We here collect all the conditions we need to impose in order to stabilize the system around the de Sitter solutions against
anisotropy, ghosts and gradient instabilities.
As the attractor condition for the background dynamics, we impose
\begin{equation}
\vert C_{\rm att} \vert \ll 1 \; ,
\label{Catt-all}
\end{equation}
and take the limit $C_{\rm att} \rightarrow 0$ for the pure de Sitter. Also for the background evolution, the condition to prevent from rolling to anisotropic configuration away from \eqref{A-vev} is
\begin{equation}
C_{\rm iso} \ge 0 \; ,
\label{Ciso-all}
\end{equation}
where $C_{\rm iso}$ is defined in \eqref{Ciso}.

Regarding perturbations, we are concerned with instabilities in the high momentum limit on the background (\ref{dS-H}--\ref{EQBGdS2}).
As noted below \eqref{CNG7}, $C_{\rm NG7} \approx C_{\rm NG3}$ in the (quasi) de Sitter limit; moreover, $C_{\rm NG4}$ can be expressed as
\begin{equation}
C_{\rm NG4} = \frac{C_{\rm NG2}}{2} + \frac{C_{\rm NG5}}{2 C_{\rm NG1}} \left( C_{\rm NG2} + 2 G_{4,X}^2 B^2 \right) \; ,
\end{equation}
and hence $C_{\rm NG4} > 0$ is granted as long as the other no-ghost conditions are respected.
Therefore in order to ensure stability against ghosts, it suffice to impose the following $5$ inequalities
\begin{equation}
C_{\rm NG1} > 0 \; , \quad C_{\rm NG2} > 0 \; , \quad C_{\rm NG3} > 0 \; , \quad C_{\rm NG5} > 0 \; , \quad C_{\rm NG6} > 0 \; ,
\label{noG-all}
\end{equation}
where the definitions can be found in \eqref{CNG12}, (\ref{CNG3}--\ref{CNG5}) and \eqref{CNG6}.
One immediate observation is that since $C_{\rm NG1} + C_{\rm NG3} = 2 ( \bar G_{2, W_1} + 2 \bar G_{2,W_2} )B^2 > 0$ (remember to take $\dot B \approx \dot H \approx 0$ in the expression \eqref{CNG3} for $C_{\rm NG3}$), one must have a model with $\bar G_{2,W_1} \ne 0$ or $\bar G_{2,W_2} \ne 0$, at the very least, for a no-ghost inflationary solution. This is a quick evidence of ghosts in the existing models of vector-driven inflation.

On the other hand, the gradient stabilities demand, in the same large $k$ and de Sitter limit,
\begin{equation}
\begin{aligned}
& \alpha_T > 0 \; , \quad \beta_T > 0 \; , \quad \alpha_T^2 - 4 \beta_T \ge 0 \; , \\
& \alpha_V > 0 \; , \quad \beta_V > 0 \; , \quad \alpha_V^2 - 4 \beta_V \ge 0 \; , \\
& \gamma_S > 0 \; , \quad  \delta_S > 0 \; , \quad \epsilon_S > 0 \; , \quad \gamma_S^2 - 3 \, \delta_S \ge 0 \; , \quad
\gamma_S^2 \delta_S^2 - 4 \delta_S^3 - 4 \gamma_S^3 \epsilon_S + 18 \gamma_S \delta_S \epsilon_S - 27 \epsilon_S^2 \ge 0 \; ,
\end{aligned}
\label{gradstab-all}
\end{equation}
where $T,V,S$ correspond to the tensor, vector and scalar sectors, respectively. While the expressions for $\alpha_T$ and $\beta_T$ are given in \eqref{gradstab-T1} and \eqref{gradstab-T2}, respectively, those for $\alpha_V, \beta_V, \gamma_S,\delta_S,\epsilon_S$ are to be shown and discussed with concrete examples in the next section.
For a stable inflationary solution to be realized in the class of models \eqref{action}, all of the conditions in \eqref{noG-all} and \eqref{gradstab-all} need to be satisfied at all times.

\section{Illustrative examples}
\label{sec:examples}

In the previous section, we have analyzed the stability around the (quasi) de Sitter background (\ref{dS-H}--\ref{EQBGdS2}) in the theory \eqref{action} without specifying any functional forms of $G_2$ or $G_4$.
In this section, we consider a few concrete examples. In particular, first we take a few existing, therefore unstable, cases in Sec.~\ref{subsec:example1} and demonstrate how ghost and/or gradient instabilities appear. Then we provide a simple yet successful example that evades all the stability conditions \eqref{noG-all} and \eqref{gradstab-all} in Sec.~\ref{subsec:example2}.
We also impose the condition under which the background is an attractor, i.e.~$\vert C_{\rm att} \vert \ll 1$ with $C_{\rm att}$ given in \eqref{cond-attractor}, and its de-Sitter limit by $C_{\rm att} \rightarrow 0$.

\subsection{Existing (unstable) examples}
\label{subsec:example1}

Existing models in the literature in which inflation is driven solely by  vector fields without the non-Abelian-specific structure $\epsilon_{abc} \partial_{[\mu}A_{\nu]}^{(a)} A^{(b)\mu} A^{(c)\nu}$ can be classified into three main categories: (i) potential driven one \cite{Ford:1989me}, (ii) non-minimal coupling \cite{Turner:1987bw,Golovnev:2008cf,Dimopoulos:2008rf}, and (iii) fixed norm of $A_\mu A^\mu$ \cite{Ackerman:2007nb}. All of these models have been found unstable, by ghost and/or gradient instabilities \cite{Himmetoglu:2008hx,Himmetoglu:2008zp,Himmetoglu:2009qi}. Here we translate their results to the ones in our framework and illustrate the cause of instabilities.
The first two can actually be combined with the choice of functions $G_2 = Y - V(X)$ and $G_4 = M_p^2 / 2 + \xi X$ with some constant $\xi$, so we study them together.%
\footnote{In the original works \cite{Golovnev:2008cf,Dimopoulos:2008rf}, $\xi = - 1 / 6$ was taken, but we leave it arbitrary in our analysis.}
One note to make is that in \cite{Turner:1987bw,Golovnev:2008cf,Dimopoulos:2008rf}, the term proportional to $G_{4,X}$ in \eqref{def-Ls} is set to be zero by hand, but this is expected to lead to a higher-derivative ghost, or Ostrogradsky instability, at the nonlinear level, and here we include it for consistency and show that instabilities appear nonetheless.%
\footnote{This higher-derivative ghost mode is not excited at the order of quadratic action, as is evident from the full analysis in \cite{Himmetoglu:2009qi}. The reason is a technical one: this ghost is associated with the longitudinal mode $A_\mu^L = \partial_\mu \chi$ and arises from the term $A^2 R \supset - 2 A_0^2 \, \partial_t K / N^3$, where $N$ is the lapse and $K$ is the trace of the extrinsic curvature $K_{ij} \supset \partial_t g_{ij} / 2N$. Since $\langle A_0 \rangle = 0$ on the background, the ghost becomes relevant only for $A^2 R \propto ( \partial_t \chi )^2 \partial_t \phi$, where $\phi \equiv - \delta N$, and after $\phi$ is integrated out, with our gauge fixing choice. Thus this should occur in the cubic action or higher.}
The third category fixes the value of $A^2$ by the term of Lagrange multiplier $\lambda (A^2 - M^2)$, driving inflation; however due to the presence of an additional variable, $\lambda$, this model does not fall in the domain of the class of models characterized by \eqref{action}.
Therefore, we retract our concern from this model (the presence of instabilities is studied in detail in \cite{Himmetoglu:2008hx}) and focus on the first two models combined.

The model in question is with a potential term of the vector fields and their non-minimal coupling to gravity, motivated by \cite{Turner:1987bw,Ford:1989me,Golovnev:2008cf,Dimopoulos:2008rf}. Although these original works did not include the counter terms proportional to $G_{4,X}$ in \eqref{def-Ls} that eliminates a non-linear, higher-derivative ghost, we do so for a consistent treatment. The original model is known to suffer from instabilities \cite{Himmetoglu:2008zp,Himmetoglu:2009qi} already, and we here demonstrate that the same conclusion holds even with the inclusion of the counter terms.
This model is characterized by the choice of functions
\begin{equation}
G_2 = Y - V(X) \; , \quad G_4 = \frac{M_p^2}{2} + \xi X \; ,
\end{equation}
with a constant $\xi$.
The background equations in de Sitter, (\ref{EQBGdS1}--\ref{EQBGdS2}), read
\begin{equation}
3 M_p^2 H^2 \approx \frac{3 \left( 1 + 6 \xi \right)}{2} \, H^2 B^2 + V \; , \quad
2 \left( 1 + 3 \xi \right) H^2 \approx V_{,X} \; ,
\end{equation}
as well as $\dot B \approx \dot H \approx 0$.
In order for this to be an attractor, one needs \eqref{Catt-all}
\begin{equation}
\vert C_{\rm att} \vert = \frac{B^2}{H^2} \left\vert 3 V_{,XX} - \frac{4 \left( 1 + 3 \xi \right) H^2 + 3 \left( 1+ 4 \xi \right)^2 V_{,XX} B^2}{4 C_{\rm NG6}} \right\vert \ll 1 \; .
\label{attractor-ex1}
\end{equation}
The stability against anisotropic expansion requires \eqref{Ciso-all}
\begin{equation}
C_{\rm iso} = \frac{6\xi - 1}{4} \ge 0 \; ,
\end{equation}
and therefore $\xi \ge 1/6$ is necessary.

The no-ghost conditions \eqref{noG-all} in this model are satisfied by making the following quantities positive:
\begin{equation}
C_{\rm NG1} = C_{\rm NG5} = 1 \; , \quad
C_{\rm NG2} = \frac{M_p^2}{2} + \frac{\xi \left( 1 - 4 \xi \right)}{2} B^2 \; , \quad
C_{\rm NG3} = -1 \; , \quad
C_{\rm NG6} = C_{\rm NG2} + 6 \xi^2 B^2 \; .
\label{noG-ex1}
\end{equation}
Since $C_{\rm NG3} = -1$, one vector and one scalar modes are always ghosty, at least near de Sitter.
On the other hand, the squared sound speeds for tensor and vector modes, $c_s^T$ and $c_s^V$ respectively, can be solved easily, yielding
\begin{eqnarray}
c_s^{T \,2} & = & 1 \, , \;
1 + \frac{\xi B^2}{C_{\rm NG2}} \; ,
\label{cT2-ex1} \\
c_s^{V \, 2} & = & 1 \, , \;
1 + \frac{\xi^2 B^2}{C_{\rm NG2}} \; .
\label{cV2-ex1}
\end{eqnarray}
Also the squared sound speed of one of the scalar modes is also unity. The other two can be obtained by solving the equation $( c_s^{S \, 2} )^2 - \alpha_S \, c_s^{S\, 2} + \beta_S = 0$, where
\begin{equation}
\alpha_S = 2 - \frac{V_{, XX} B^2}{2 H^2} - \frac{2 \xi^3 \left( 1 - 2 \xi \right)}{C_{\rm NG2} C_{\rm NG6}} B^4 \; , \quad
\beta_S = \left( 1 - \frac{V_{, XX} B^2}{2 H^2} \right) \left[ 1 - \frac{2 \xi^3 \left( 1 - 4 \xi \right)}{C_{\rm NG2} C_{\rm NG6}} B^4 \right] \; ,
\label{cS2-ex1}
\end{equation}
and their positivity and reality can be ensured by requiring $\alpha_S > 0$, $\beta_S > 0$ and $\alpha_S^2 - 4 \beta_S > 0$ simultaneously.

By looking at (\ref{attractor-ex1}--\ref{cS2-ex1}), one can see that the conditions for no gradient instabilities around attractor solutions can be achieved in, for example, the cases where $B^2 \ll M_p^2$ and $V_{,XX} B^2 \ll H^2$ with reasonable values of $\xi$; however, the system suffer ghosts in the vector and scalar sectors. Note also that in the case $B^2 \gg M_p^2$ with $V \propto X$ and $\xi = -1/6$, we additionally have negative values of $C_{\rm iso}$, $C_{\rm NG2}$, $C_{\rm NG4}$, $\alpha_S$ and $\beta_S$, indicating anisotropic background expansion, ghosts in all the sectors and gradient instabilities in the scalar sector. Therefore, this model with vector potential and non-minimal coupling to Ricci scalar is always unstable around inflationary backgrounds.

\subsection{A simple successful example}
\label{subsec:example2}

We have so far considered known models of vector-driven inflation and identified the causes of instabilities. In this subsection, we provide a proof of existence by demonstrating a simple example that can satisfy all of attractor, no-ghost and gradient stability conditions simultaneously.
First we turn off $Z$ (vanishing on the background) and $W_3$, while we need to preserve $G_{2,W_1}$ or $G_{2,W_2}$ for ghost stability, as mentioned below \eqref{noG-all}, and thus we do both. We then consider a class of models that contain terms linear in $W_{1,2}$. We postulate the following simple forms of $G_2$ and $G_4$ to be
\begin{equation}
G_2 = F(X,Y) + c_1 W_1 + c_2 W_2 \; , \quad
G_4 = c_3 + c_4 X \; ,
\label{model-success}
\end{equation}
where $c_i$ are constants and $F$ is a function of $X$ and $Y$.
To further simplify our analysis, we consider the case where $F$ has properties
\begin{equation}
\bar F_{,XX} + H^2 \bar F_{,XY} =
H^2 \left( H^2 \bar F_{,YY} + \bar F_{,XY} \right)
= 0 \; ,
\end{equation}
where bar denotes quantities on the quasi de Sitter background, which is given by the solutions to (\ref{dS-H}--\ref{EQBGdS2}), i.e.~$\dot H \approx \dot B \approx 0$ and
\begin{equation}
6 c_3 \approx - \frac{\bar F}{H^2} + 3 \bar F_{,Y} B^2 - 3 \left( c_1 + 3 c_2 \right) B^4 + 9 c_4 B^2 \; , \quad
\frac{\bar F_{,X}}{H^2} + 2 \bar F_{,Y} - 2 \left( c_1 + 3 c_2 \right) B^2 + 6 c_4 \approx 0 \; .
\label{EQBG-stableEX}
\end{equation}
For the attractor condition $\vert C_{\rm att} \vert \ll 1$, the expression \eqref{cond-attractor} now reduces to
\begin{equation}
C_{\rm att} = - \frac{B^2}{2} \, \frac{
4 \left[ \bar F_{,Y}^2 + 3 c_4 \bar F_{,Y} + 4 \left( c_1 + 3 c_2 \right) c_3 \right]
+ 3 \left[ 5 \bar F_{,Y} B^2 - 2 \left( c_1 + 3 c_2 \right) B^4 + 8 c_3 + 12 c_4 B^2 \right] \bar F_{,XY}}
{\left( 2 c_3 + c_4 B^2 \right) \left[ \bar F_{,Y} - \left( 3 \bar F_{,XY} + 2 c_1 + 6 c_2 \right) B^2 \right] + 8 c_4^2 B^2} \; .
\end{equation}
The isotropic configuration described by (\ref{A-vev}) and (\ref{FLRW}) is preserved if \eqref{Ciso-all}, i.e.
\begin{equation}
C_{\rm iso} = \frac{3}{2} \left( c_1 + 2 c_2 \right) \left[ 2 \left( c_1 + 3c_2 \right) B^2 - 3 c_4 \right] B^2
- c_1 c_3
- \frac{1}{4} \, \bar F_{,Y}^2
+ \frac{3}{2} \left( c_4 - c_2 B^2 \right) \bar F_{,Y}
\ge 0 \; ,
\label{Ciso-EXsimple}
\end{equation}
is satisfied.
The parameters for the no-ghost conditions are
\begin{equation}
\begin{aligned}
& C_{\rm NG1} = C_{\rm NG5} = \bar F_{,Y} - 2 c_1 B^2 \; , \quad
C_{\rm NG2} = C_{\rm NG1} \left( c_3 + \frac{c_4}{2} B^2 \right) - 2 c_4^2 B^2 \; , \quad
C_{\rm NG3} = - \bar F_{,Y} + 4 \left( c_1 + c_2 \right) B^2 \; , \\
& C_{\rm NG6} = \frac{1}{2} \, \frac{\left( 2 c_3 + c_4 B^2 \right) \left[ \bar F_{,Y} - \left( 3 \bar F_{,XY} + 2 c_1 + 6 c_2 \right) B^2 \right] + 8 c_4^2 B^2}{\bar F_{,Y} - \left( \bar F_{,XY} + 2 c_1 + 2 c_2 \right) B^2} \; ,
\end{aligned}
\end{equation}
and we require $C_{\rm NG1}, C_{\rm NG2}, C_{\rm NG3}, C_{\rm NG6} > 0$ simultaneously. Among the parameters in this model, we can express $\bar F_{,Y}$, $\bar F_{,XY}$, $c_1$ and $c_3$ in terms of $C_{\rm NG1}$, $C_{\rm NG2}$, $C_{\rm NG3}$ and $C_{\rm NG6}$. We replace them in the expressions of other stability conditions, and they are now all written in terms of $c_2, c_4, C_{\rm NG1}, C_{\rm NG2}, C_{\rm NG3}, C_{\rm NG6}$ and $B$, which can now be treated as independent model parameters (note that $\bar F$ and $\bar F_{,X}$ are eliminated by using the background equations \eqref{EQBG-stableEX}).
Further, by considering the de-Sitter limit of the attractor, we take $C_{\rm att} \rightarrow 0$, with which we can replace $C_{\rm NG6}$. Then all the stability conditions are expressed by the independent parameters $c_2, c_4, C_{\rm NG1}, C_{\rm NG2}, C_{\rm NG3}$ and $B$.
To illustrate the stable region in the parameter space, we expand those expressions in the limit $\vert c_4 \vert \gg \vert c_2 \vert B^2 \gg C_{\rm NG1,3}, C_{\rm NG2} / B^2$, finding
\begin{eqnarray}
\alpha_T & = & \frac{ 8 \, c_2 c_4^2 B^4}{C_{\rm NG1} C_{\rm NG2}} + {\cal O} \left( c_4^1, c_2^0 \right) \; , \quad
\beta_T = \frac{ 8 \, c_2 c_4^2 B^4}{C_{\rm NG1} C_{\rm NG2}} + {\cal O} \left( c_4^1, c_2^0 \right) \; , \\
\alpha_V & = & \frac{ 8 \, c_2^2 c_4^2 B^6}{C_{\rm NG1} C_{\rm NG2} C_{\rm NG3}} + {\cal O} \left( c_4^1, c_2^1 \right) \; , \quad
\beta_V = \frac{ 8 \, c_2^3 c_4^2 B^8}{C_{\rm NG1}^2 C_{\rm NG2} C_{\rm NG3}} + {\cal O} \left( c_4^1, c_2^2 \right) \\
\gamma_S & = & \frac{32 \, c_2^2 c_4^2 B^6}{3 \, C_{\rm NG1} C_{\rm NG2} C_{\rm NG3}} + {\cal O} \left( c_4^1 , c_2^1 \right) \; , \quad
\delta_S = \frac{128 \, c_2^3 c_4^2 B^8}{3 \, C_{\rm NG1}^2 C_{\rm NG2} C_{\rm NG3}} + {\cal O} \left( c_4^1 , c_2^2 \right) \; , \\
\epsilon_S & = & \frac{28 \left( C_{\rm NG1} - C_{\rm NG3} \right) c_2^2 c_4^2 B^6}{3\, C_{\rm NG1}^2 C_{\rm NG2} C_{\rm NG3}} + {\cal O} \left( c_4^1 , c_2^1 \right) \; .
\end{eqnarray}
Hence we can immediately see that all the stability conditions can be satisfied when $c_2 > 0$ and $C_{\rm NG1} > C_{\rm NG3}$, provided that $\vert c_4 \vert \gg c_2 B^2 \gg C_{\rm NG1,3}, C_{\rm NG2} / B^2$.
Notice that the model has sufficient number of independent parameters to achieve this set of conditions.
The reality conditions of $c_s^2$ are also fulfilled, as $\alpha_T^2 - 4 \beta_T \approx \alpha_T^2 > 0$, $\alpha_V^2 - 4 \beta_V \approx \alpha_V^2 > 0$, $\gamma_S^2 - 3 \, \delta_S \approx \gamma_S^2 >  0$, and $\gamma_S^2 \delta_S^2 - 4 \delta_S^3 - 4 \gamma_S^3 \epsilon_S + 18 \gamma_S \delta_S \epsilon_S - 27 \epsilon_S^2 \approx \gamma_S^2 \delta_S^2 > 0$.
Moreover, in the same limit, we have, from \eqref{Ciso-EXsimple},
\begin{equation}
C_{\rm iso} = \frac{4 \, c_2 c_4^2 B^2}{C_{\rm NG1}} + {\cal O} \left( c_4^1, c_2^0 \right)
\end{equation}
and thus the isotropic condition is also satisfied.

Albeit some required hierarchy and tuning of parameters, we have presented a simple model \eqref{model-success} that has an inflationary solution as an isotropic attractor fulfilling (\ref{Catt-all}) and (\ref{Ciso-all}) and that satisfies all the stability conditions \eqref{noG-all} and \eqref{gradstab-all}.
This is, to our knowledge, the first example of stable inflationary models in which inflation is driven only by vector fields without relying on the terms specific to non-Abelian gauge fields.

\section{Conclusion}
\label{sec:conclusion}

We have demonstrated that, in a new class of models where vector fields are solely responsible for inflation, quasi de-Sitter solutions can be an inflationary attractor that is stable against all the pathological instabilities.
It has been known that, in most of vector-driven inflationary models, violation of gauge invariance is invoked in order to sustain a sufficiently long period of accelerated expansion, but it leads to propagating longitudinal modes of the vector fields that suffer from ghost and/or linear instabilities on the desired background. While some non-Abelian gauge field models are known to provide stable inflationary backgrounds, spectra of perturbations in those models are incompatible with observational data. This implies that the gauge symmetry should be broken in those models. Once the gauge symmetry is broken, a more general class of models is allowed/motivated.
In this paper, we have explored a new class of model Lagrangians that allows for stable inflationary solutions driven solely by vector fields.

In order to perform a general analysis and then to narrow down the stability conditions, we have taken the Generalized Proca theory with an additional global $O(3)$ symmetry as our starting point, see \eqref{action}. Quasi de-Sitter solutions for the background, $\dot{H} \approx 0$, are then sought for, and the ``slow roll'' of the vector vev, $\dot{B} \approx 0$, immediately follows, unless there is some fine-tuning of the model functions, as discussed right after \eqref{dS-B}. For these solutions to be an isotropic inflationary attractor, two conditions, \eqref{Catt-all} and \eqref{Ciso-all}, are demanded, and we have provided their explicit expressions in \eqref{cond-attractor} and \eqref{Ciso}.
Proceeding the analysis of linearized perturbations, we have decomposed them into $3$ decoupled sectors, ``tensor,'' ``vector'' and ``scalar,'' where the decomposition is done with respect to the vector-field internal space as well as to the spatial coordinates. We have outlined the general procedure to obtain the conditions to evade both the ghost and gradient instabilities and applied it to each sector separately.
The set of no-ghost conditions makes it visible that the simultaneous fulfillment of them requires (not necessarily suffices) to have $G_2$ depend on at least one of $W_1 = A_\mu^{(a)} A_\nu^{(a)} F^{(b) \mu\rho} F^{(b)\nu}{}_\rho$ and $W_2 = A_\mu^{(a)} A_\nu^{(b)} F^{(a) \mu\rho} F^{(b)\nu}{}_\rho$. Within the class of model Lagrangian (\ref{action})--(\ref{def-XYZW}), these terms are therefore necessary ingredients for a vector model with inflationary solutions without ghosts.

We have obtained the attractor, no-ghost and gradient-stability conditions for the general case. In order to show that there exists a model that has a parameter window to satisfy all those conditions, we have chosen simple examples. This also helps to make the analysis of gradient stability more illuminating, since the conditions are quite lengthy in expression for the general case, especially those of vector and scalar perturbations.
Two examples are considered: the first one, specified by $G_2 = Y - V(X)$ and $G_4 = M_p^2 / 2 + \xi X$, is motivated by a few existing, unstable models, and the second is a new, successful model without instabilities, given by $G_2 = F(X,Y) + c_1 W_1 + c_2 W_2$ and $G_4 = c_3 + c_4 X$.
The first example resembles the combination of the vector field model with a potential and the one with non-minimal coupling to gravity, while the difference is the inclusion of the terms proportional to $G_{4,X}$ in the action \eqref{action}, those that eliminate the Ostrogradsky ghost degree of freedom in the nonlinear orders. It is shown that despite the inclusion of these counter terms, the model has no parameter region to have stable inflationary solutions.
The second example is truly stable model -- at least against ghost and gradient instabilities in the high momentum limit -- and we have explicitly shown that it admits a viable scenario to respect all of the attractor and stability conditions.

Our aim of this paper is devoted to the demonstration of viable models of inflation that are healthy at the theoretical level, which has turned out successful. Phenomenological consequences of such pathology-free models are therefore of interest for further studies.
The observational bounds on the scalar spectral index and the tensor-to-scalar ratio can constrain/falsify these models. In the simple example we have considered, there is no helicity dependence in the tensor sector, but for models with $G_{2,Z} \ne 0$ on the background, where $Z = - F_{\mu\nu}^{(a)} \tilde{F}^{(a)\mu\nu} /4$, parity is broken, and the left- and right-handed tensor modes are expected to be produced by different amounts.
Also, the vector perturbations are dynamical, as opposed to scalar-tensor theories, and their fate potentially leaves distinctive signatures in the CMB/LSS and other observable spectra.
Moreover, various non-standard interaction terms should arise at cubic and higher orders, and thus higher-order correlation functions are expected to have interesting features.
In addition, due to the unique behavior of the scalar, vector and tensor modes in this model, we would expect that the CMB \cite{Dai:2013ikl} and the large scale clustering fossils, \cite{Jeong:2012df, Dai:2013kra, Dimastrogiovanni:2014ina, Dimastrogiovanni:2015pla}, coming from this model would be distinguishable from the ordinary realizations of the scalar, vector and tensor modes.
Cross correlations, especially parity-odd correlators \cite{Lue:1998mq}, may also provide a unique arena to test the models of this class (notice that parity violation exists even without the presence of $Z$ terms, as the vector mode $U_a$ in fact behaves as a pseudo vector).

While the primary interest of the present paper is in inflationary models of the early universe, the quasi de Sitter solutions that we have found may provide the origin of the accelerated expansion of the universe today as well. In this case the attractor condition, $|C_{\rm att}|\ll 1$ for the inflationary attractor, is replaced by $C_{\rm att}\gtrsim 0$ and any non-negative or slightly-negative value of $C_{\rm att}$ is allowed. Compared with the previous model of dark energy based on the generalized Proca field~\cite{DeFelice:2016yws,DeFelice:2016uil}, one of the distinguishing features of the current model is the existence of extra tensor modes. This may have some impacts on gravitational-wave physics. While astrophysical objects such as binary systems excite the ordinary gravitational wave modes $h_{ij}$, they are not mass eigenstates but are linear combinations of them (see the off-diagonal component of (\ref{eqn:tensor-mixing})). Mass eigenstates with different masses then propagate differently. Thus the gravitational waves are expected to exhibit oscillations with the other tensor modes $t_{ia}$ as they propagate towards gravitational wave detectors. This implies that the detectors will observe linear combinations that are different from those at the source. This kind of phenomenon analogous to neutrino oscillation has been considered in the context of bimetric theories of gravity in the literature~\cite{DeFelice:2013nba}. The quasi de Sitter solutions found in the present paper provides a new theoretical basis for the gravitational wave oscillation.

\acknowledgments

R.E. and Y.Z. are very much grateful to both of Kavli IPMU and YITP for their invitation as well as their very warm hospitality during the initial stage of this work and their later collaboration. The work of R.E.  was supported by Hong Kong University through the
CRF Grants of the Government of the Hong Kong SAR under HKUST4/CRF/13. R.N. was supported in part by the Natural Sciences and Engineering
Research Council (NSERC) of Canada and by the Lorne Trottier Chair in Astrophysics and Cosmology at McGill, as well as by the World Premier International Research Center Initiative (WPI), MEXT, Japan. The work of SM was supported by Japan Society for the Promotion of Science (JSPS) Grants-in-Aid for Scientific Research (KAKENHI) No. 24540256, and by World Premier International Research Center Initiative (WPI), MEXT, Japan. The work of YZ was supported by the NSFC grant No. 11605228, the China Postdoctoral Science Foundation Grant No. 2016M590134, MEXT KAKENHI Grant Nos. 15H05888 and 15K21733.



\begin{thebibliography}{999}


\bibitem{Gasperini:2002bn}
  M.~Gasperini and G.~Veneziano,
  Phys.\ Rept.\  {\bf 373}, 1 (2003)
  doi:10.1016/S0370-1573(02)00389-7
  [hep-th/0207130].


\bibitem{Brandenberger:1988aj}
  R.~H.~Brandenberger and C.~Vafa,
  Nucl.\ Phys.\ B {\bf 316}, 391 (1989).
  doi:10.1016/0550-3213(89)90037-0


\bibitem{Wands:1998yp}
  D.~Wands,
  Phys.\ Rev.\ D {\bf 60}, 023507 (1999)
  doi:10.1103/PhysRevD.60.023507
  [gr-qc/9809062].


\bibitem{Finelli:2001sr}
  F.~Finelli and R.~Brandenberger,
  Phys.\ Rev.\ D {\bf 65}, 103522 (2002)
  doi:10.1103/PhysRevD.65.103522
  [hep-th/0112249].


\bibitem{Khoury:2001wf}
  J.~Khoury, B.~A.~Ovrut, P.~J.~Steinhardt and N.~Turok,
  Phys.\ Rev.\ D {\bf 64}, 123522 (2001)
  doi:10.1103/PhysRevD.64.123522
  [hep-th/0103239].


\bibitem{Buchbinder:2007ad}
  E.~I.~Buchbinder, J.~Khoury and B.~A.~Ovrut,
  Phys.\ Rev.\ D {\bf 76}, 123503 (2007)
  doi:10.1103/PhysRevD.76.123503
  [hep-th/0702154].


\bibitem{Horava:2009uw}
  P.~Horava,
  Phys.\ Rev.\ D {\bf 79}, 084008 (2009)
  doi:10.1103/PhysRevD.79.084008
  [arXiv:0901.3775 [hep-th]].


\bibitem{Mukohyama:2009gg}
  S.~Mukohyama,
  JCAP {\bf 0906}, 001 (2009)
  doi:10.1088/1475-7516/2009/06/001
  [arXiv:0904.2190 [hep-th]].


\bibitem{Creminelli:2010ba}
  P.~Creminelli, A.~Nicolis and E.~Trincherini,
  JCAP {\bf 1011}, 021 (2010)
  doi:10.1088/1475-7516/2010/11/021
  [arXiv:1007.0027 [hep-th]].


\bibitem{Bennett:1996ce}
  C.~L.~Bennett {\it et al.},
  Astrophys.\ J.\  {\bf 464}, L1 (1996)
  doi:10.1086/310075
  [astro-ph/9601067].


  G.~Hinshaw {\it et al.} [WMAP Collaboration],
  Astrophys.\ J.\ Suppl.\  {\bf 208}, 19 (2013)
  doi:10.1088/0067-0049/208/2/19
  [arXiv:1212.5226 [astro-ph.CO]].


  A.~Kogut {\it et al.},
  JCAP {\bf 1107}, 025 (2011)
  doi:10.1088/1475-7516/2011/07/025
  [arXiv:1105.2044 [astro-ph.CO]].


  T.~Matsumura {\it et al.},
  J.\ Low.\ Temp.\ Phys.\  {\bf 176}, 733 (2014)
  doi:10.1007/s10909-013-0996-1
  [arXiv:1311.2847 [astro-ph.IM]].


\bibitem{Ade:2015lrj}
  P.~A.~R.~Ade {\it et al.} [Planck Collaboration],
  Astron.\ Astrophys.\  {\bf 594}, A20 (2016)
  doi:10.1051/0004-6361/201525898
  [arXiv:1502.02114 [astro-ph.CO]].


\bibitem{Netterfield:2001yq}
  C.~B.~Netterfield {\it et al.} [Boomerang Collaboration],
  Astrophys.\ J.\  {\bf 571}, 604 (2002)
  doi:10.1086/340118
  [astro-ph/0104460].


  B.~Reichborn-Kjennerud {\it et al.},
  Proc.\ SPIE Int.\ Soc.\ Opt.\ Eng.\  {\bf 7741}, 1C (2010)
  doi:10.1117/12.857138
  [arXiv:1007.3672 [astro-ph.CO]].


  A.~A.~Fraisse {\it et al.},
  JCAP {\bf 1304}, 047 (2013)
  doi:10.1088/1475-7516/2013/04/047
  [arXiv:1106.3087 [astro-ph.CO]].


\bibitem{Carlstrom:2009um}
  J.~E.~Carlstrom {\it et al.},
  Publ.\ Astron.\ Soc.\ Pac.\  {\bf 123}, 568 (2011)
  doi:10.1086/659879
  [arXiv:0907.4445 [astro-ph.IM]].


  J.~E.~Austermann {\it et al.},
  Proc.\ SPIE Int.\ Soc.\ Opt.\ Eng.\  {\bf 8452}, 84521E (2012)
  doi:10.1117/12.927286
  [arXiv:1210.4970 [astro-ph.IM]].


  J.~L.~Sievers {\it et al.} [Atacama Cosmology Telescope Collaboration],
  JCAP {\bf 1310}, 060 (2013)
  doi:10.1088/1475-7516/2013/10/060
  [arXiv:1301.0824 [astro-ph.CO]].


  Z.~Ahmed {\it et al.} [BICEP3 Collaboration],
  Proc.\ SPIE Int.\ Soc.\ Opt.\ Eng.\  {\bf 9153}, 91531N (2014)
  doi:10.1117/12.2057224
  [arXiv:1407.5928 [astro-ph.IM]].


  T.~Essinger-Hileman {\it et al.},
  Proc.\ SPIE Int.\ Soc.\ Opt.\ Eng.\  {\bf 9153}, 91531I (2014)
  doi:10.1117/12.2056701
  [arXiv:1408.4788 [astro-ph.IM]].


  B.~A.~Benson {\it et al.} [SPT-3G Collaboration],
  Proc.\ SPIE Int.\ Soc.\ Opt.\ Eng.\  {\bf 9153}, 91531P (2014)
  doi:10.1117/12.2057305
  [arXiv:1407.2973 [astro-ph.IM]].


  P.~A.~R.~Ade {\it et al.} [BICEP2 and Planck Collaborations],
  Phys.\ Rev.\ Lett.\  {\bf 114}, 101301 (2015)
  doi:10.1103/PhysRevLett.114.101301
  [arXiv:1502.00612 [astro-ph.CO]].


  T.~Louis {\it et al.},
  arXiv:1610.02360 [astro-ph.CO].


\bibitem{Martin:2013tda}
  J.~Martin, C.~Ringeval and V.~Vennin,
  Phys.\ Dark Univ.\  {\bf 5-6}, 75 (2014)
  doi:10.1016/j.dark.2014.01.003
  [arXiv:1303.3787 [astro-ph.CO]].


\bibitem{Martin:2013nzq}
  J.~Martin, C.~Ringeval, R.~Trotta and V.~Vennin,
  JCAP {\bf 1403}, 039 (2014)
  doi:10.1088/1475-7516/2014/03/039
  [arXiv:1312.3529 [astro-ph.CO]].


\bibitem{ArmendarizPicon:1999rj}
  C.~Armendariz-Picon, T.~Damour and V.~F.~Mukhanov,
  Phys.\ Lett.\ B {\bf 458}, 209 (1999)
  doi:10.1016/S0370-2693(99)00603-6
  [hep-th/9904075].


\bibitem{Kaloper:1991rw}
  N.~Kaloper,
  Phys.\ Rev.\ D {\bf 44}, 2380 (1991).
  doi:10.1103/PhysRevD.44.2380


  E.~Di Grezia, G.~Esposito, A.~Funel, G.~Mangano and G.~Miele,
  Phys.\ Rev.\ D {\bf 68}, 105012 (2003)
  doi:10.1103/PhysRevD.68.105012
  [gr-qc/0305050].


  T.~S.~Koivisto and N.~J.~Nunes,
  Phys.\ Lett.\ B {\bf 685}, 105 (2010)
  doi:10.1016/j.physletb.2010.01.051
  [arXiv:0907.3883 [astro-ph.CO]].


  T.~S.~Koivisto and N.~J.~Nunes,
  Phys.\ Rev.\ D {\bf 80}, 103509 (2009)
  doi:10.1103/PhysRevD.80.103509
  [arXiv:0908.0920 [astro-ph.CO]].


  T.~Kobayashi and S.~Yokoyama,
  JCAP {\bf 0905}, 004 (2009)
  doi:10.1088/1475-7516/2009/05/004
  [arXiv:0903.2769 [astro-ph.CO]].


  C.~Germani and A.~Kehagias,
  JCAP {\bf 0903}, 028 (2009)
  doi:10.1088/1475-7516/2009/03/028
  [arXiv:0902.3667 [astro-ph.CO]].


  T.~S.~Koivisto, D.~F.~Mota and C.~Pitrou,
  JHEP {\bf 0909}, 092 (2009)
  doi:10.1088/1126-6708/2009/09/092
  [arXiv:0903.4158 [astro-ph.CO]].


  F.~R.~Urban,
  JCAP {\bf 1308}, 008 (2013)
  doi:10.1088/1475-7516/2013/08/008
  [arXiv:1306.6429 [astro-ph.CO]].


  K.~S.~Kumar, J.~Marto, N.~J.~Nunes and P.~V.~Moniz,
  JCAP {\bf 1406}, 064 (2014)
  doi:10.1088/1475-7516/2014/06/064
  [arXiv:1404.0211 [gr-qc]].


  B.~J.~Barros and N.~J.~Nunes,
  Phys.\ Rev.\ D {\bf 93}, no. 4, 043512 (2016)
  doi:10.1103/PhysRevD.93.043512
  [arXiv:1511.07856 [astro-ph.CO]].


\bibitem{Turner:1987bw}
  M.~S.~Turner and L.~M.~Widrow,
  Phys.\ Rev.\ D {\bf 37}, 2743 (1988).
  doi:10.1103/PhysRevD.37.2743


\bibitem{Golovnev:2008cf}
  A.~Golovnev, V.~Mukhanov and V.~Vanchurin,
  JCAP {\bf 0806}, 009 (2008)
  doi:10.1088/1475-7516/2008/06/009
  [arXiv:0802.2068 [astro-ph]].


\bibitem{Kanno:2008gn}
  S.~Kanno, M.~Kimura, J.~Soda and S.~Yokoyama,
  JCAP {\bf 0808}, 034 (2008)
  doi:10.1088/1475-7516/2008/08/034
  [arXiv:0806.2422 [hep-ph]].


\bibitem{Dimopoulos:2008rf}
  K.~Dimopoulos and M.~Karciauskas,
  JHEP {\bf 0807}, 119 (2008)
  doi:10.1088/1126-6708/2008/07/119
  [arXiv:0803.3041 [hep-th]].


\bibitem{Ford:1989me}
  L.~H.~Ford,
  Phys.\ Rev.\ D {\bf 40}, 967 (1989).
  doi:10.1103/PhysRevD.40.967


\bibitem{Koivisto:2008xf}
  T.~Koivisto and D.~F.~Mota,
  JCAP {\bf 0808}, 021 (2008)
  doi:10.1088/1475-7516/2008/08/021
  [arXiv:0805.4229 [astro-ph]].


\bibitem{Ackerman:2007nb}
  L.~Ackerman, S.~M.~Carroll and M.~B.~Wise,
  Phys.\ Rev.\ D {\bf 75}, 083502 (2007)
  Erratum: [Phys.\ Rev.\ D {\bf 80}, 069901 (2009)]
  doi:10.1103/PhysRevD.75.083502, 10.1103/PhysRevD.80.069901
  [astro-ph/0701357].


\bibitem{Himmetoglu:2008hx}
  B.~Himmetoglu, C.~R.~Contaldi and M.~Peloso,
  Phys.\ Rev.\ D {\bf 79}, 063517 (2009)
  doi:10.1103/PhysRevD.79.063517
  [arXiv:0812.1231 [astro-ph]].


\bibitem{Himmetoglu:2008zp}
  B.~Himmetoglu, C.~R.~Contaldi and M.~Peloso,
  Phys.\ Rev.\ Lett.\  {\bf 102}, 111301 (2009)
  doi:10.1103/PhysRevLett.102.111301
  [arXiv:0809.2779 [astro-ph]].


\bibitem{Himmetoglu:2009qi}
  B.~Himmetoglu, C.~R.~Contaldi and M.~Peloso,
  Phys.\ Rev.\ D {\bf 80}, 123530 (2009)
  doi:10.1103/PhysRevD.80.123530
  [arXiv:0909.3524 [astro-ph.CO]].


\bibitem{Watanabe:2009ct}
  M.~a.~Watanabe, S.~Kanno and J.~Soda,
  Phys.\ Rev.\ Lett.\  {\bf 102}, 191302 (2009)
  doi:10.1103/PhysRevLett.102.191302
  [arXiv:0902.2833 [hep-th]].


\bibitem{Ratra:1991bn}
  B.~Ratra,
  Astrophys.\ J.\  {\bf 391}, L1 (1992).
  doi:10.1086/186384


\bibitem{Bamba:2003av}
  K.~Bamba and J.~Yokoyama,
  Phys.\ Rev.\ D {\bf 69}, 043507 (2004)
  doi:10.1103/PhysRevD.69.043507
  [astro-ph/0310824].


\bibitem{Martin:2007ue}
  J.~Martin and J.~Yokoyama,
  JCAP {\bf 0801}, 025 (2008)
  doi:10.1088/1475-7516/2008/01/025
  [arXiv:0711.4307 [astro-ph]].


\bibitem{Demozzi:2009fu}
  V.~Demozzi, V.~Mukhanov and H.~Rubinstein,
  JCAP {\bf 0908}, 025 (2009)
  doi:10.1088/1475-7516/2009/08/025
  [arXiv:0907.1030 [astro-ph.CO]].

\bibitem{Emami:2009vd}
R.~Emami, H.~Firouzjahi and M.~S.~Movahed,
Phys.\ Rev.\ D {\bf 81}, 083526 (2010)
doi:10.1103/PhysRevD.81.083526
[arXiv:0908.4161 [hep-th]].

\bibitem{Caldwell:2011ra}
R.~R.~Caldwell, L.~Motta and M.~Kamionkowski,
Phys.\ Rev.\ D {\bf 84}, 123525 (2011)
doi:10.1103/PhysRevD.84.123525
[arXiv:1109.4415 [astro-ph.CO]].


\bibitem{Fujita:2012rb}
  T.~Fujita and S.~Mukohyama,
  JCAP {\bf 1210}, 034 (2012)
  doi:10.1088/1475-7516/2012/10/034
  [arXiv:1205.5031 [astro-ph.CO]].


\bibitem{Ferreira:2013sqa}
  R.~J.~Z.~Ferreira, R.~K.~Jain and M.~S.~Sloth,
  JCAP {\bf 1310}, 004 (2013)
  doi:10.1088/1475-7516/2013/10/004
  [arXiv:1305.7151 [astro-ph.CO]].


\bibitem{Kobayashi:2014sga}
  T.~Kobayashi,
  JCAP {\bf 1405}, 040 (2014)
  doi:10.1088/1475-7516/2014/05/040
  [arXiv:1403.5168 [astro-ph.CO]].


\bibitem{Ferreira:2014mwa}
  R.~Z.~Ferreira and J.~Ganc,
  JCAP {\bf 1504}, no. 04, 029 (2015)
  doi:10.1088/1475-7516/2015/04/029
  [arXiv:1411.5362 [astro-ph.CO]].


\bibitem{Domenech:2015zzi}
  G.~Domènech, C.~Lin and M.~Sasaki,
  Europhys.\ Lett.\  {\bf 115}, no. 1, 19001 (2016)
  doi:10.1209/0295-5075/115/19001
  [arXiv:1512.01108 [astro-ph.CO]].


\bibitem{Campanelli:2015jfa}
  L.~Campanelli,
  Eur.\ Phys.\ J.\ C {\bf 75}, no. 6, 278 (2015)
  doi:10.1140/epjc/s10052-015-3510-x
  [arXiv:1503.07415 [gr-qc]].


\bibitem{Fujita:2016qab}
  T.~Fujita and R.~Namba,
  Phys.\ Rev.\ D {\bf 94}, no. 4, 043523 (2016)
  doi:10.1103/PhysRevD.94.043523
  [arXiv:1602.05673 [astro-ph.CO]].


\bibitem{Barnaby:2012tk}
  N.~Barnaby, R.~Namba and M.~Peloso,
  Phys.\ Rev.\ D {\bf 85}, 123523 (2012)
  doi:10.1103/PhysRevD.85.123523
  [arXiv:1202.1469 [astro-ph.CO]].


\bibitem{Fujita:2013pgp}
  T.~Fujita and S.~Yokoyama,
  JCAP {\bf 1309}, 009 (2013)
  doi:10.1088/1475-7516/2013/09/009
  [arXiv:1306.2992 [astro-ph.CO]].


\bibitem{Fujita:2014sna}
  T.~Fujita and S.~Yokoyama,
  JCAP {\bf 1403}, 013 (2014)
  Erratum: [JCAP {\bf 1405}, E02 (2014)]
  doi:10.1088/1475-7516/2014/03/013, 10.1088/1475-7516/2014/05/E02
  [arXiv:1402.0596 [astro-ph.CO]].


\bibitem{Tasinato:2014fia}
  G.~Tasinato,
  JCAP {\bf 1503}, 040 (2015)
  doi:10.1088/1475-7516/2015/03/040
  [arXiv:1411.2803 [hep-th]].


\bibitem{Mukohyama:2016npi}
  S.~Mukohyama,
  arXiv:1607.07041 [hep-th].


\bibitem{Anber:2009ua}
  M.~M.~Anber and L.~Sorbo,
  Phys.\ Rev.\ D {\bf 81}, 043534 (2010)
  doi:10.1103/PhysRevD.81.043534
  [arXiv:0908.4089 [hep-th]].


\bibitem{Barnaby:2010vf}
  N.~Barnaby and M.~Peloso,
  Phys.\ Rev.\ Lett.\  {\bf 106}, 181301 (2011)
  doi:10.1103/PhysRevLett.106.181301
  [arXiv:1011.1500 [hep-ph]].


\bibitem{Barnaby:2011vw}
  N.~Barnaby, R.~Namba and M.~Peloso,
  JCAP {\bf 1104}, 009 (2011)
  doi:10.1088/1475-7516/2011/04/009
  [arXiv:1102.4333 [astro-ph.CO]].


\bibitem{Cook:2011hg}
  J.~L.~Cook and L.~Sorbo,
  Phys.\ Rev.\ D {\bf 85}, 023534 (2012)
  Erratum: [Phys.\ Rev.\ D {\bf 86}, 069901 (2012)]
  doi:10.1103/PhysRevD.86.069901, 10.1103/PhysRevD.85.023534
  [arXiv:1109.0022 [astro-ph.CO]].


\bibitem{Barnaby:2011qe}
  N.~Barnaby, E.~Pajer and M.~Peloso,
  Phys.\ Rev.\ D {\bf 85}, 023525 (2012)
  doi:10.1103/PhysRevD.85.023525
  [arXiv:1110.3327 [astro-ph.CO]].


\bibitem{Crowder:2012ik}
  S.~G.~Crowder, R.~Namba, V.~Mandic, S.~Mukohyama and M.~Peloso,
  Phys.\ Lett.\ B {\bf 726}, 66 (2013)
  doi:10.1016/j.physletb.2013.08.077
  [arXiv:1212.4165 [astro-ph.CO]].


\bibitem{Namba:2015gja}
  R.~Namba, M.~Peloso, M.~Shiraishi, L.~Sorbo and C.~Unal,
  JCAP {\bf 1601}, no. 01, 041 (2016)
  doi:10.1088/1475-7516/2016/01/041
  [arXiv:1509.07521 [astro-ph.CO]].

\bibitem{Obata:2016tmo}
  I.~Obata {\it et al.} [CLEO Collaboration],
  Phys.\ Rev.\ D {\bf 93}, no. 12, 123502 (2016)
  doi:10.1103/PhysRevD.93.123502
  [arXiv:1602.06024 [hep-th]].

\bibitem{Maleknejad:2016qjz}
  A.~Maleknejad,
  JHEP {\bf 1607}, 104 (2016)
  doi:10.1007/JHEP07(2016)104
  [arXiv:1604.03327 [hep-ph]].

\bibitem{Obata:2016oym}
  I.~Obata,
  arXiv:1612.08817 [astro-ph.CO].


\bibitem{Caprini:2014mja}
  C.~Caprini and L.~Sorbo,
  JCAP {\bf 1410}, no. 10, 056 (2014)
  doi:10.1088/1475-7516/2014/10/056
  [arXiv:1407.2809 [astro-ph.CO]].


\bibitem{Fujita:2015iga}
  T.~Fujita, R.~Namba, Y.~Tada, N.~Takeda and H.~Tashiro,
  JCAP {\bf 1505}, no. 05, 054 (2015)
  doi:10.1088/1475-7516/2015/05/054
  [arXiv:1503.05802 [astro-ph.CO]].


\bibitem{Adshead:2016iae}
  P.~Adshead, J.~T.~Giblin, T.~R.~Scully and E.~I.~Sfakianakis,
  JCAP {\bf 1610}, 039 (2016)
  doi:10.1088/1475-7516/2016/10/039
  [arXiv:1606.08474 [astro-ph.CO]].


\bibitem{Anber:2015yca}
  M.~M.~Anber and E.~Sabancilar,
  Phys.\ Rev.\ D {\bf 92}, no. 10, 101501 (2015)
  doi:10.1103/PhysRevD.92.101501
  [arXiv:1507.00744 [hep-th]].


\bibitem{Cado:2016kdp}
  Y.~Cado and E.~Sabancilar,
  arXiv:1611.02293 [hep-ph].


\bibitem{Linde:2012bt}
  A.~Linde, S.~Mooij and E.~Pajer,
  Phys.\ Rev.\ D {\bf 87}, no. 10, 103506 (2013)
  doi:10.1103/PhysRevD.87.103506
  [arXiv:1212.1693 [hep-th]].


\bibitem{McDonough:2016xvu}
  E.~McDonough, H.~Bazrafshan Moghaddam and R.~H.~Brandenberger,
  JCAP {\bf 1605}, no. 05, 012 (2016)
  doi:10.1088/1475-7516/2016/05/012
  [arXiv:1601.07749 [hep-th]].


\bibitem{Pajer:2013fsa}
  E.~Pajer and M.~Peloso,
  Class.\ Quant.\ Grav.\  {\bf 30}, 214002 (2013)
  doi:10.1088/0264-9381/30/21/214002
  [arXiv:1305.3557 [hep-th]].


\bibitem{Lue:1998mq}
A.~Lue, L.~M.~Wang and M.~Kamionkowski,
Phys.\ Rev.\ Lett.\  {\bf 83}, 1506 (1999)
doi:10.1103/PhysRevLett.83.1506
[astro-ph/9812088].


\bibitem{Adshead:2012kp}
  P.~Adshead and M.~Wyman,
  Phys.\ Rev.\ Lett.\  {\bf 108}, 261302 (2012)
  doi:10.1103/PhysRevLett.108.261302
  [arXiv:1202.2366 [hep-th]].



\bibitem{Maleknejad:2011jw}
  A.~Maleknejad and M.~M.~Sheikh-Jabbari,
  Phys.\ Lett.\ B {\bf 723}, 224 (2013)
  doi:10.1016/j.physletb.2013.05.001
  [arXiv:1102.1513 [hep-ph]].


\bibitem{Maleknejad:2011sq}
  A.~Maleknejad and M.~M.~Sheikh-Jabbari,
  Phys.\ Rev.\ D {\bf 84}, 043515 (2011)
  doi:10.1103/PhysRevD.84.043515
  [arXiv:1102.1932 [hep-ph]].


\bibitem{SheikhJabbari:2012qf}
  M.~M.~Sheikh-Jabbari,
  Phys.\ Lett.\ B {\bf 717}, 6 (2012)
  doi:10.1016/j.physletb.2012.09.014
  [arXiv:1203.2265 [hep-th]].


\bibitem{Adshead:2012qe}
  P.~Adshead and M.~Wyman,
  Phys.\ Rev.\ D {\bf 86}, 043530 (2012)
  doi:10.1103/PhysRevD.86.043530
  [arXiv:1203.2264 [hep-th]].


\bibitem{Maleknejad:2011jr}
  A.~Maleknejad, M.~M.~Sheikh-Jabbari and J.~Soda,
  JCAP {\bf 1201}, 016 (2012)
  doi:10.1088/1475-7516/2012/01/016
  [arXiv:1109.5573 [hep-th]].


\bibitem{Dimastrogiovanni:2012ew}
  E.~Dimastrogiovanni and M.~Peloso,
  Phys.\ Rev.\ D {\bf 87}, no. 10, 103501 (2013)
  doi:10.1103/PhysRevD.87.103501
  [arXiv:1212.5184 [astro-ph.CO]].


\bibitem{Adshead:2013nka}
  P.~Adshead, E.~Martinec and M.~Wyman,
  JHEP {\bf 1309}, 087 (2013)
  doi:10.1007/JHEP09(2013)087
  [arXiv:1305.2930 [hep-th]].


\bibitem{Namba:2013kia}
  R.~Namba, E.~Dimastrogiovanni and M.~Peloso,
  JCAP {\bf 1311}, 045 (2013)
  doi:10.1088/1475-7516/2013/11/045
  [arXiv:1308.1366 [astro-ph.CO]].


\bibitem{Nieto:2016gnp}
  C.~M.~Nieto and Y.~Rodriguez,
  Mod.\ Phys.\ Lett.\ A {\bf 31}, no. 21, 1640005 (2016)
  doi:10.1142/S0217732316400058
  [arXiv:1602.07197 [gr-qc]].


\bibitem{Adshead:2016omu}
  P.~Adshead, E.~Martinec, E.~I.~Sfakianakis and M.~Wyman,
  arXiv:1609.04025 [hep-th].

\bibitem{Tasinato:2014eka}
  G.~Tasinato,
  JHEP {\bf 1404}, 067 (2014)
  doi:10.1007/JHEP04(2014)067
  [arXiv:1402.6450 [hep-th]].


\bibitem{Heisenberg:2014rta}
  L.~Heisenberg,
  JCAP {\bf 1405}, 015 (2014)
  doi:10.1088/1475-7516/2014/05/015
  [arXiv:1402.7026 [hep-th]].


\bibitem{Allys:2015sht}
  E.~Allys, P.~Peter and Y.~Rodriguez,
  JCAP {\bf 1602}, no. 02, 004 (2016)
  doi:10.1088/1475-7516/2016/02/004
  [arXiv:1511.03101 [hep-th]].


\bibitem{Jimenez:2016isa}
  J.~Beltran Jimenez and L.~Heisenberg,
  Phys.\ Lett.\ B {\bf 757}, 405 (2016)
  doi:10.1016/j.physletb.2016.04.017
  [arXiv:1602.03410 [hep-th]].

\bibitem{Allys:2016jaq}
   E.~Allys, J.~P.~Beltran Almeida, P.~Peter and Y.~Rodr\'{i}guez,
   JCAP {\bf 1609} (2016) no.09,  026
   doi:10.1088/1475-7516/2016/09/026
   [arXiv:1605.08355 [hep-th]].


\bibitem{Ostrogradsky:1850fid}
  M.~Ostrogradsky,
  Mem.\ Acad.\ St.\ Petersbourg {\bf 6}, no. 4, 385 (1850).


\bibitem{Heisenberg:2016eld}
  L.~Heisenberg, R.~Kase and S.~Tsujikawa,
  Phys.\ Lett.\ B {\bf 760}, 617 (2016)
  doi:10.1016/j.physletb.2016.07.052
  [arXiv:1605.05565 [hep-th]].

\bibitem{Kimura:2016rzw}
  R.~Kimura, A.~Naruko and D.~Yoshida,
  JCAP {\bf 1701}, no. 01, 002 (2017)
  doi:10.1088/1475-7516/2017/01/002
  [arXiv:1608.07066 [gr-qc]].


\bibitem{Oliveros:2016myr}
  A.~Oliveros,
  doi:10.1007/s10509-016-2998-3
  arXiv:1612.06377 [gr-qc].


\bibitem{DeFelice:2016yws}
  A.~De Felice, L.~Heisenberg, R.~Kase, S.~Mukohyama, S.~Tsujikawa and Y.~l.~Zhang,
  JCAP {\bf 1606}, no. 06, 048 (2016)
  doi:10.1088/1475-7516/2016/06/048
  [arXiv:1603.05806 [gr-qc]].

\bibitem{DeFelice:2016uil}
  A.~De Felice, L.~Heisenberg, R.~Kase, S.~Mukohyama, S.~Tsujikawa and Y.~l.~Zhang,
  Phys.\ Rev.\ D {\bf 94}, no. 4, 044024 (2016)
  doi:10.1103/PhysRevD.94.044024
  [arXiv:1605.05066 [gr-qc]].


\bibitem{Watanabe:2010fh}
  M.~a.~Watanabe, S.~Kanno and J.~Soda,
  Prog.\ Theor.\ Phys.\  {\bf 123}, 1041 (2010)
  doi:10.1143/PTP.123.1041
  [arXiv:1003.0056 [astro-ph.CO]].


\bibitem{Gumrukcuoglu:2010yc}
  A.~E.~Gumrukcuoglu, B.~Himmetoglu and M.~Peloso,
  Phys.\ Rev.\ D {\bf 81}, 063528 (2010)
  doi:10.1103/PhysRevD.81.063528
  [arXiv:1001.4088 [astro-ph.CO]].


\bibitem{Emami:2010rm}
R.~Emami, H.~Firouzjahi, S.~M.~Sadegh Movahed and M.~Zarei,
JCAP {\bf 1102}, 005 (2011)
doi:10.1088/1475-7516/2011/02/005
[arXiv:1010.5495 [astro-ph.CO]].


\bibitem{Emami:2011yi}
  R.~Emami and H.~Firouzjahi,
  JCAP {\bf 1201}, 022 (2012)
  doi:10.1088/1475-7516/2012/01/022
  [arXiv:1111.1919 [astro-ph.CO]].


\bibitem{Soda:2012zm}
  J.~Soda,
  Class.\ Quant.\ Grav.\  {\bf 29}, 083001 (2012)
  doi:10.1088/0264-9381/29/8/083001
  [arXiv:1201.6434 [hep-th]].


\bibitem{Shiraishi:2012xt}
  M.~Shiraishi, S.~Saga and S.~Yokoyama,
  JCAP {\bf 1211}, 046 (2012)
  doi:10.1088/1475-7516/2012/11/046
  [arXiv:1209.3384 [astro-ph.CO]].


\bibitem{Bartolo:2012sd}
  N.~Bartolo, S.~Matarrese, M.~Peloso and A.~Ricciardone,
  Phys.\ Rev.\ D {\bf 87}, no. 2, 023504 (2013)
  doi:10.1103/PhysRevD.87.023504
  [arXiv:1210.3257 [astro-ph.CO]].


\bibitem{Emami:2013bk}
  R.~Emami and H.~Firouzjahi,
  JCAP {\bf 1310}, 041 (2013)
  doi:10.1088/1475-7516/2013/10/041
  [arXiv:1301.1219 [hep-th]].


\bibitem{Shiraishi:2013vja}
  M.~Shiraishi, E.~Komatsu, M.~Peloso and N.~Barnaby,
  JCAP {\bf 1305}, 002 (2013)
  doi:10.1088/1475-7516/2013/05/002
  [arXiv:1302.3056 [astro-ph.CO]].


\bibitem{Abolhasani:2013zya}
  A.~A.~Abolhasani, R.~Emami, J.~T.~Firouzjaee and H.~Firouzjahi,
  JCAP {\bf 1308}, 016 (2013)
  doi:10.1088/1475-7516/2013/08/016
  [arXiv:1302.6986 [astro-ph.CO]].

\bibitem{Lyth:2013kah}
  D.~H.~Lyth and M.~Karciauskas,
  JCAP {\bf 1305}, 011 (2013)
  doi:10.1088/1475-7516/2013/05/011
  [arXiv:1302.7304 [astro-ph.CO]].


\bibitem{Baghram:2013lxa}
  S.~Baghram, M.~H.~Namjoo and H.~Firouzjahi,
  JCAP {\bf 1308}, 048 (2013)
  doi:10.1088/1475-7516/2013/08/048
  [arXiv:1303.4368 [astro-ph.CO]].


\bibitem{Biagetti:2013qqa}
  M.~Biagetti, A.~Kehagias, E.~Morgante, H.~Perrier and A.~Riotto,
  JCAP {\bf 1307}, 030 (2013)
  doi:10.1088/1475-7516/2013/07/030
  [arXiv:1304.7785 [astro-ph.CO]].


\bibitem{Ohashi:2013qba}
  J.~Ohashi, J.~Soda and S.~Tsujikawa,
  JCAP {\bf 1312}, 009 (2013)
  doi:10.1088/1475-7516/2013/12/009
  [arXiv:1308.4488 [astro-ph.CO], arXiv:1308.4488].


\bibitem{Abolhasani:2013bpa}
  A.~A.~Abolhasani, R.~Emami and H.~Firouzjahi,
  JCAP {\bf 1405}, 016 (2014)
  doi:10.1088/1475-7516/2014/05/016
  [arXiv:1311.0493 [hep-th]].


\bibitem{Ramazanov:2013wea}
  S.~R.~Ramazanov and G.~Rubtsov,
  Phys.\ Rev.\ D {\bf 89}, no. 4, 043517 (2014)
  doi:10.1103/PhysRevD.89.043517
  [arXiv:1311.3272 [astro-ph.CO]].


\bibitem{Chen:2014eua}
X.~Chen, R.~Emami, H.~Firouzjahi and Y.~Wang,
JCAP {\bf 1408}, 027 (2014)
doi:10.1088/1475-7516/2014/08/027
[arXiv:1404.4083 [astro-ph.CO]].


\bibitem{Naruko:2014bxa}
  A.~Naruko, E.~Komatsu and M.~Yamaguchi,
  JCAP {\bf 1504}, no. 04, 045 (2015)
  doi:10.1088/1475-7516/2015/04/045
  [arXiv:1411.5489 [astro-ph.CO]].


\bibitem{Emami:2015uva}
  R.~Emami and H.~Firouzjahi,
  JCAP {\bf 1510}, no. 10, 043 (2015)
  doi:10.1088/1475-7516/2015/10/043
  [arXiv:1506.00958 [astro-ph.CO]].


\bibitem{Emami:2015qjl}
R.~Emami,
arXiv:1511.01683 [astro-ph.CO].


\bibitem{Abolhasani:2015cve}
A.~A.~Abolhasani, M.~Akhshik, R.~Emami and H.~Firouzjahi,
JCAP {\bf 1603}, no. 03, 020 (2016)
doi:10.1088/1475-7516/2016/03/020
[arXiv:1511.03218 [astro-ph.CO]].


\bibitem{Dimopoulos:2008yv}
  K.~Dimopoulos, M.~Karciauskas, D.~H.~Lyth and Y.~Rodriguez,
  JCAP {\bf 0905}, 013 (2009)
  doi:10.1088/1475-7516/2009/05/013
  [arXiv:0809.1055 [astro-ph]].


\bibitem{Dimopoulos:2009am}
  K.~Dimopoulos, M.~Karciauskas and J.~M.~Wagstaff,
  Phys.\ Rev.\ D {\bf 81}, 023522 (2010)
  doi:10.1103/PhysRevD.81.023522
  [arXiv:0907.1838 [hep-ph]].


\bibitem{Dimopoulos:2009vu}
  K.~Dimopoulos, M.~Karciauskas and J.~M.~Wagstaff,
  Phys.\ Lett.\ B {\bf 683}, 298 (2010)
  doi:10.1016/j.physletb.2009.12.024
  [arXiv:0909.0475 [hep-ph]].


\bibitem{Dimopoulos:2011ws}
  K.~Dimopoulos,
  Int.\ J.\ Mod.\ Phys.\ D {\bf 21}, 1250023 (2012)
  Erratum: [Int.\ J.\ Mod.\ Phys.\ D {\bf 21}, 1292003 (2012)]
  doi:10.1142/S021827181250023X, 10.1142/S0218271812920034
  [arXiv:1107.2779 [hep-ph]].


\bibitem{Namba:2012gg}
  R.~Namba,
  Phys.\ Rev.\ D {\bf 86}, 083518 (2012)
  doi:10.1103/PhysRevD.86.083518
  [arXiv:1207.5547 [astro-ph.CO]].

\bibitem{Heisenberg:2016wtr}
  L.~Heisenberg, R.~Kase and S.~Tsujikawa,
  JCAP {\bf 1611}, no. 11, 008 (2016)
  doi:10.1088/1475-7516/2016/11/008
  [arXiv:1607.03175 [gr-qc]].


\bibitem{Kim:2013gka}
  J.~Kim and E.~Komatsu,
  Phys.\ Rev.\ D {\bf 88}, 101301 (2013)
  doi:10.1103/PhysRevD.88.101301
  [arXiv:1310.1605 [astro-ph.CO]].


\bibitem{Jimenez:2016upj}
  J.~B.~Jiménez and L.~Heisenberg,
  arXiv:1610.08960 [hep-th].


\bibitem{Horndeski:1974wa}
  G.~W.~Horndeski,
  Int.\ J.\ Theor.\ Phys.\  {\bf 10}, 363 (1974).
  doi:10.1007/BF01807638


\bibitem{Nicolis:2008in}
  A.~Nicolis, R.~Rattazzi and E.~Trincherini,
  Phys.\ Rev.\ D {\bf 79}, 064036 (2009)
  doi:10.1103/PhysRevD.79.064036
  [arXiv:0811.2197 [hep-th]].


\bibitem{Deffayet:2011gz}
  C.~Deffayet, X.~Gao, D.~A.~Steer and G.~Zahariade,
  Phys.\ Rev.\ D {\bf 84}, 064039 (2011)
  doi:10.1103/PhysRevD.84.064039
  [arXiv:1103.3260 [hep-th]].


\bibitem{Deffayet:2009wt}
  C.~Deffayet, G.~Esposito-Farese and A.~Vikman,
  Phys.\ Rev.\ D {\bf 79}, 084003 (2009)
  doi:10.1103/PhysRevD.79.084003
  [arXiv:0901.1314 [hep-th]].


\bibitem{Kobayashi:2011nu}
  T.~Kobayashi, M.~Yamaguchi and J.~Yokoyama,
  Prog.\ Theor.\ Phys.\  {\bf 126}, 511 (2011)
  doi:10.1143/PTP.126.511
  [arXiv:1105.5723 [hep-th]].


\bibitem{Gumrukcuoglu:2016jbh}
  A.~E.~G\"{u}mr\"{u}k\c{c}\"{u}o\v{g}lu, S.~Mukohyama and T.~P.~Sotiriou,
  Phys.\ Rev.\ D {\bf 94}, no. 6, 064001 (2016)
  doi:10.1103/PhysRevD.94.064001
  [arXiv:1606.00618 [hep-th]].

\bibitem{Dai:2013ikl}
L.~Dai, D.~Jeong and M.~Kamionkowski,
Phys.\ Rev.\ D {\bf 87}, no. 10, 103006 (2013)
doi:10.1103/PhysRevD.87.103006
[arXiv:1302.1868 [astro-ph.CO]].

\bibitem{Dai:2013kra}
L.~Dai, D.~Jeong and M.~Kamionkowski,
Phys.\ Rev.\ D {\bf 88}, no. 4, 043507 (2013)
doi:10.1103/PhysRevD.88.043507
[arXiv:1306.3985 [astro-ph.CO]].

\bibitem{Jeong:2012df}
D.~Jeong and M.~Kamionkowski,
Phys.\ Rev.\ Lett.\  {\bf 108}, 251301 (2012)
doi:10.1103/PhysRevLett.108.251301
[arXiv:1203.0302 [astro-ph.CO]].


\bibitem{Dimastrogiovanni:2014ina}
E.~Dimastrogiovanni, M.~Fasiello, D.~Jeong and M.~Kamionkowski,
JCAP {\bf 1412}, 050 (2014)
doi:10.1088/1475-7516/2014/12/050
[arXiv:1407.8204 [astro-ph.CO]].

\bibitem{Dimastrogiovanni:2015pla}
E.~Dimastrogiovanni, M.~Fasiello and M.~Kamionkowski,
JCAP {\bf 1602}, 017 (2016)
doi:10.1088/1475-7516/2016/02/017
[arXiv:1504.05993 [astro-ph.CO]].




\bibitem{DeFelice:2013nba}
  A.~De Felice, T.~Nakamura and T.~Tanaka,
  PTEP {\bf 2014}, 043E01 (2014)
  doi:10.1093/ptep/ptu024
  [arXiv:1304.3920 [gr-qc]].

\end{thebibliography}
\end{document}